\documentclass[english,superscriptaddress,twocolumn,oneside, amsmath,amssymb,amsfonts,aps,pra,floatfix]{revtex4-1}
\usepackage[english]{babel}
\usepackage[utf8x]{inputenc}
\usepackage{microtype}
\usepackage{graphicx}
\usepackage{siunitx}
\usepackage{xspace}
\usepackage[hidelinks]{hyperref}
\usepackage{bm}
\usepackage{bbold}
\usepackage{natbib}
\usepackage{csquotes}
\usepackage{xcolor}
\usepackage{qcircuit} 

\usepackage[draft]{fixme}

\usepackage[capitalise]{cleveref}

\newcommand{\ketbra}[2]{\ensuremath{\left| #1 \right\rangle\!\!\left\langle #2 \right|}}
\newcommand{\unity}{\ensuremath{\mathbb{1}}}
\newcommand{\ket}[1]{\ensuremath{| #1 \nobreak \rangle } }
\crefname{section}{Sec.}{Sec.}

\newcommand{\affA}{Department of Physics and Astronomy, Aarhus University, DK-8000 Aarhus C, Denmark}
\newcommand{\affB}{Aarhus Institute of Advanced Studies, Aarhus University, DK-8000 Aarhus C, Denmark}
\newcommand{\affUvA}{Institute of Physics, University of Amsterdam, Science Park 904, 1098 XH Amsterdam, the Netherlands}
\newcommand{\affQuSoft}{QuSoft, Science Park 123, 1098 XG Amsterdam, the Netherlands}

\date{\today}
\begin{document}
	
	\title{Single-step implementation of high-fidelity $n$-bit Toffoli gates}
	
	\author{S. E. Rasmussen}
	\email{stig@phys.au.dk}
	\affiliation{\affA}
	
	\author{K. Groenland}
	\affiliation{\affUvA}
	\affiliation{\affQuSoft}
	
	\author{R. Gerritsma}
	\affiliation{\affUvA}
	
	\author{K. Schoutens}
	\affiliation{\affUvA}
	\affiliation{\affQuSoft}

	\author{N. T. Zinner}
	\email{zinner@phys.au.dk}
	\affiliation{\affA}
	\affiliation{\affB}
	
	\begin{abstract}
		The family of $n$-bit Toffoli gates, with the two-bit Toffoli gate as the figurehead, are of great interest in quantum information as they can be used as universal gates and in quantum error correction, among other things. 
		We present a single-step implementation of arbitrary $n$-bit Toffoli gates (up to a local change of basis), based on resonantly driving a single qubit that has a strong Ising coupling to $n$ other qubits. The setup in the two-qubit case turns out to be identical to the universal Barenco gate.
		The gate time and error are, in theory, independent of the number of control qubits, scaling better than conventional circuit decompositions. We note that our assumptions, namely strongly coupling $n+1$ qubits and a driving frequency that scales with $n$, may break down for large systems. 
		Still, our protocol could enhance the capabilities of intermediate scale quantum computers, and we discuss the prospects of implementing our protocol on trapped ions, Rydberg atoms, and on superconducting circuits. Simulations of the latter platform show that the Toffoli gate with two control bits attains fidelities of above 0.98 even in the presence of decoherence.
		We also show how similar ideas can be used to make a series of controlled-\textsc{not}-gates in a single step. We show how these can speed up the implementation of quantum error correcting codes and we simulate the encoding steps of the three-qubit bit-flip code and the seven-qubit Steane code. 
	\end{abstract}
	
	\maketitle
	
	\section{Introduction}\label{sec:intro}
	
	The $n$-bit Toffoli gates are a family of reversible logic gates, where each gate has $n$ control bits and one bit which is inverted if the control bits are in the right state. The $n$-bit Toffoli gates, and especially the two-bit Toffoli gate, simply known as \emph{the} Toffoli gate \cite{Toffoli1980}, are of great interest in the field of quantum information \cite{Nielsen2010}. The two-bit Toffoli gate, on its own, is a universal gate in classical computing and together with the Hadamard gate it constitutes a universal set of quantum gates \cite{Nielsen2010}. The $n$-bit Toffoli gates are further important since they play a pivotal role in schemes for quantum error correction \cite{Cory1998,Schindler2011}, in fault-tolerant quantum computing \cite{Dennis2001,Paetznick2013}, and in Shor's algorithm \cite{Shor1995}.

	
	While high-fidelity quantum gates on one or two qubits have been reported  \cite{Ballance:2016,Gaebler:2016,Jau:2016,Maller:2015,Barends:2014}, accurate implementations of multiqubit gates such as the Toffoli gate remain challenging. 
	In a conventional circuit decomposition, where the Toffoli gate is performed as a sequence of one- and two-qubit gates, it is known that at least 5 two-qubit operations are needed to obtain a two-bit Toffoli gate. For larger $n$, these numbers grow steeply: The $n$-bit Toffoli can be implemented with a circuit of depth $O(\log(n))$, requiring $O(n)$ ancilla bits. If no ancillas may be used, the number of controlled-\textsc{not} (\textsc{cnot}) gates is lower bounded at $2n$, although the best known implementations require a quadratic number of \textsc{cnot} gates \cite{Shende2009}.
	
	Circumventing this decomposition has also attracted significant attention. Reference \cite{Ralph2007} considers a shorter circuit for the two-bit Toffoli gate by requiring one qutrit and Refs. \cite{Fedorov2012,Reed2012,Cao2018} implement a similar scheme, employing superconducting transmon qubits or atoms in coupled cavities. Other proposals rely on the properties of resonant driving, such as the two-bit Toffoli gate using a modified Jaynes-Cummings model \cite{Chen2012} or other multiqubit gates in integrable spin chains \cite{Groenland2018,Groenland2019}.
	References \cite{Isenhower2011,Molmer2011,Shi2018} describe a Toffoli gate for general $n$ by exploiting the Rydberg blockade, and Ref. \cite{Wang2001} proposes the same gate using trapped ions. Another proposal for the two-bit Toffoli gate using the Rydberg atom is based on Stark-tuned three-body Förster resonances \cite{Beterov2018}. A recent result in Ref. \cite{Gullans2019} addresses a driven two-bit Toffoli gate for silicon spin qubits.
	
	Here we present a simple single-step implementation of the $n$-bit Toffoli gate for an arbitrary $n$. We require a strong, Ising-type coupling between a `target' qubit and $n$ `control' qubits, and then apply a driving field to selectively invert the target qubit. This results in an operation we call $i$-Toffoli, which can be straightforwardly mapped into a conventional Toffoli gate by demoting a single qubit to an ancilla. Surprisingly, we find that the gate time and error do not increase with $n$ in theory, which beats previously known results. We critically note that our assumptions may break down at larger system sizes: We require an interaction between $n$ qubits and a single target, where the interaction strength should not decrease with $n$. Moreover, the required driving frequency scales with the number of qubits. Still, the protocol could greatly enhance the capabilities of certain near-term quantum computers \cite{Preskill2018}, and we perform a detailed study of its performance on superconducting circuits. Our simulations find that when decoherence is neglected the fidelity is approximately constant above 0.995, and when decoherence is included the $i$-Toffoli attains fidelities above 0.98 with up to five control qubits, for gate times of $\SI{50}{\nano\s}$.
	A similar driving approach allows a `fanout' gate, where a \textsc{cnot} gate takes place between a single control and $n$ target qubits. We discuss its application in error correction, where qubits can be encoded in fewer steps.
	
	Our proposal is closely related to previous work on multiqubit gates that exploit the Rydberg blockade interaction, espectially Ref. \cite{Isenhower2011}. In contrast to such prior art, we do not assume a perfect blockade interaction, but consider an Ising model with finite interaction strength, allowing a rigorous analysis of gate times and errors. Moreover, our broader perspective results in the same operation in fewer steps, without restricting the study to a single platform.

	The paper is organized as follows.
	In \cref{sec:cnnot} we present a simple Hamiltonian and show how it yields an $n$-bit Toffoli gate. As an example we consider the $n=1$ case, which turns out to be identical with to the universal Barenco gate. Then we discuss the gate error and the asymptotic scaling. We also discuss the effectiveness of the gate exploring the important $n=2$ case as an example in \cref{sec:example}. In \cref{sec:cnotn} we explain how to use the same ideas to implement a \textsc{cnot} gate on several qubits at the same time. We further, in \cref{sec:exerimental}, present possible implementations of the gates using superconducting circuits, Rydberg atoms, and trapped ions. In \cref{sec:QEC} we combine the gates and show how to create a more efficient quantum error correction by simulating the three-qubit bit-flip correcting code and the Steane seven-qubit code using our single step gates. In \cref{sec:conclusion} we provide a summary and outlook for future work.

	\section{Implementation of selective inversion}\label{sec:cnnot}
	
	Consider $n+1$ qubits each with frequency $\omega_j$. All of the qubits are connected with Ising coupling with strength $J_{jk}$ as described by the Ising Hamiltonian
	\begin{equation}\label{eq:Ising}
	\hat{H}_\text{Ising} = \frac{1}{2}\sum_{j<k=0}^{n} J_{jk} \sigma_j^z\sigma^z_k,
	\end{equation}
	while the non-interacting part of the Hamiltonian is given as
	\begin{equation}\label{eq:H0}
	\hat H_0 = -\frac{1}{2} \sum_{j=0}^{n}\omega_j \sigma_j^z,
	\end{equation}
	where $\sigma^{x,y,z}$ denote the Pauli operators. We denote the quantum states in the computational basis by $|x_0,\vec{x} \rangle$, where $x_0 \in \{0,1\}$ represents states of the zeroth qubit, which we will call the target qubit, and $\vec{x} \in \{0,1\}^n$ denotes the string of the state of the remaining qubits, which we call control qubits. These states are eigenstates of $H_0 + H_\text{Ising}$, whose energies we denote by $E_{x_0,\vec{x}}$.
	We drive the zeroth qubit with a field of the form
	\begin{equation}\label{eq:Hd}
	\hat H_\text{drive} = \alpha_0(t)\sigma^x_0 + \beta_0(t)\sigma^y_0.
	\end{equation}
	When the driving is included in the Hamiltonian the Hilbert space decomposes into conserved subspaces, one for each $\vec{x}$. Each of these subspaces is spanned by $|0,\vec{x}\rangle$ and $|1,\vec{x}\rangle $. We define the energy gap between such two states, due to the Ising interaction, as 
	\begin{equation}
	\begin{aligned}
	\Delta_{\vec{x}} =& E_{0,\vec{x}} - E_{1,\vec{x}} + \omega_0 \\
	=& \sum_{j=1}^{n} J_{j0}(-1)^{\vec{x}_j},
	\end{aligned}
	\end{equation}
	where $\vec{x}_j$ denotes the $j$th entry in the string of control qubit states. Similarly, we define the mean energy as $\bar{E}_{\vec{x}} = (E_{0,\vec{x}} + E_{1,\vec{x}})/2$. The Hamiltonian of a given subspace is then
	\begin{equation}
	\hat{H}_{\vec{x}} = \frac{1}{2} ( \Delta_{\vec x} - \omega_0 ) \sigma^z + \alpha(t)\sigma^x + \beta_0(t)\sigma^y + \bar E_{\vec{x}} \unity_2.
	\end{equation}
	Here $\unity_2$ denotes the two-dimensional identity matrix. 
	
	We now consider the driving fields. In general different combinations of the driving fields $\alpha_j(t)$ and $\beta_j(t)$ will lead to the same result, and here we consider a balanced two quadrature driving
	\begin{equation}\label{eq:Bi}
	\begin{aligned}
	\alpha_0(t) &= \Omega \cos\left[ ( \Delta_0 - \omega_0) t + \theta\right],\\
	\beta_0(t) &= \Omega \sin\left[ ( \Delta_0 - \omega_0 ) t + \theta\right],
	\end{aligned}
	\end{equation}
	where $\Delta_0$ is the driving frequency up to the frequency of qubit 0, $\Omega$ is the Rabi frequency, and $\theta$ is the driving phase. We now transform into the rotating frame using the transformation
	\begin{equation} \label{eq:Uint}
	\hat U_\text{int}(t) = \exp \left(i\left[\hat H_0 + \frac{1}{2}\Delta_0 \sigma_0^z + \sum_{\vec x \in \{0,1\}^n}\bar E_{\vec{x}} |\vec x \rangle \langle \vec x | \right]t\right).
	\end{equation}
	In this frame, for each subspace labeled by $\vec x$, the Hamiltonian takes the form, 
	\begin{equation}\label{eq:HI}
	\hat H_{\vec{x},I} =  \delta_{\vec {x}}\sigma^z_0 + \Omega(\sigma_0^x\cos\theta + \sigma_0^y\sin\theta),
	\end{equation}
	where $\delta_{\vec x} = (\Delta_{\vec{x}} - \Delta_0)/2$ defines the detuning. With the now time-independent Hamiltonian we can calculate the time evolution operator for all two-dimensional subspaces
	\begin{equation}\label{eq:timeEvo}
	\hat U(t) = \bigoplus_{\vec{x} \in \{ 0,1 \}^n } \left(\unity_2 \cos v_{\vec{x}}t - i \frac{\vec \sigma \cdot \vec v_{\vec{x}}}{v_{\vec{x}}}\sin v_{\vec{x}}t\right),
	\end{equation}
	where $\vec \sigma = (\sigma^x,\sigma^y,\sigma^z)$, and
	\begin{equation}
	\vec v_{\vec{x}} = \begin{bmatrix}
	\Omega \cos \theta \\
	\Omega \sin \theta \\
	\delta_{\vec{x}}
	\end{bmatrix},
	\end{equation}
	with $v_{\vec{x}} = |\vec v_{\vec{x}}|$ being the length of the vector.
	
	It follows from \cref{eq:timeEvo} that we have obtained selective state inversion. In order to see this we consider the case were the driving frequency is resonant with an energy gap of a single subspace $\vec{x}'$, i.e., $\Delta_0 = \Delta_{\vec{x}'}$, in which case we obtain a rotation around a vector in the ($x$-$y$)-plane, leading to a perfect inversion at times $T = (2m+1)\pi/2\Omega$ for $m\in \mathbb{Z}$, where the time-evolution operator of that subspace takes the form 
	\begin{equation}\label{eq:Uxnot}
	\hat U_{\vec{x}'}(t=T) = \pm i \left(\sigma^x \cos\theta + \sigma^y \sin\theta\right).
	\end{equation}
	The remaining off-resonant subspaces, i.e., assuming $|\Omega| \ll |\Delta_0 - \Delta_{\vec{x}}|$, evolve approximately as if no driving takes place:
	\begin{equation} \label{eq:UxID}
	\hat U_{\vec{x}}(t=T) \approx  \exp (-i\delta_{\vec {x}}\sigma^z t).
	\end{equation}
	Thus we conclude that if we set $\Delta_0 = \Delta_{\vec{x}'}$ we obtain an inversion of the resonant subspaces, while the off-resonant subspaces are not inverted.
	
	Note that we do not require the $J_{jk}$'s to be equal, but we do require them to be larger than the Rabi frequency, i.e., $J_{jk} \gg \Omega$, to satisfy the off-resonance condition. We further note that if, instead of the two quadratures in \cref{eq:Bi}, we had used one quadrature driving, i.e., $\beta(t) = 0$, we would have had two driving fields of opposite sign: 
	\begin{equation}
	\alpha(t) =2\Omega \left( e^{i (\Delta_0-\omega_0) t} + e^{-i (\Delta_0-\omega_0) t} \right).
	\end{equation}
	When $\omega_0 = 0$, there would be two resonant subspaces in which the zeroth qubit is inverted. This problem is fixed by demanding a relatively large frequency of the zeroth qubit, $\omega_0 \gg \Omega$. Moreover, in the case $\beta=0$ the above results are then no longer exact, but remain valid if the rotating-wave approximation ($\Delta_{\vec{x}},  \Delta_{0} \gg \Omega$) applies.
	
	\subsection{The Barenco gates}\label{sec:Barenco}
	
	In classical reversible computing there is no two-qubit gate which is both universal and reversible. However, in quantum computing any entangling two-qubit gate is universal when assisted by one-qubit gates \cite{DiVincenzo1995,Bremner2002}. Some two-qubit gates are even universal on their own. The first two-qubit gates which were shown to be universal were the family of Barenco gates \cite{Barenco1995a}, and it turns out that our implementation above yields exactly such gates for $n=1$. Therefore, and for the sake of an example, we discuss the $n=1$ more in depth.
	
	Consider the Hamiltonian $\hat H = \hat H_0 + \hat H_\text{Ising} + \hat H_\text{drive}$ for $n=1$. In this case the Hamiltonian splits up into two subspaces: $\{|00\rangle, |10\rangle\}$ and $\{|01\rangle, |11\rangle\}$. We now transform into the interacting picture using the transformation
	\begin{equation}
	\hat U_{\text{int}} = \exp \left[ \hat H_0 - \delta_1 \sigma_1^z + \frac{1}{2}\delta_1 \unity_4 + \Delta_0\sigma^z_0 \sigma^z_1  \right],
	\end{equation}
	where $\delta_1$ is some detuning from the frequency of the control qubit. Now if we require the driving to be on resonance with the target qubit, i.e., $\Delta_0 = -J_{10}$ then the interacting Hamiltonian takes the form
	\begin{equation}
	\hat H_I = \delta_1(\ketbra{01}{01} + \ketbra{11}{11}) + \Omega (\sigma^+ e^{-i\theta} + \sigma^- e^{i\theta}).
	\end{equation}
	Exponentiating this to get the time evolution operator we obtain
	\begin{equation}
	\hat U (t)= \begin{bmatrix}
	1 & 0 & 0 & 0 \\
	0 & 1 & 0 & 0 \\
	0 & 0 & e^{i\delta_1t}\cos \Omega t & -i e^{i(\delta_1t - \theta)}\sin \Omega t \\
	0 & 0 & -i e^{i(\delta_1t + \theta)}\sin \Omega t & e^{i\delta_1t}\cos \Omega t \\
	\end{bmatrix},
	\end{equation}
	which is identical to the family of Barenco gates. 
	
	As the Barenco gates are closely related to the Deutsch gate, this begs the question whether our implementation yields a Deutsch gate for $n=2$. However, this turns out not to be the case, there is a phase of $i$ to differ.
	
	\subsection{The ($n-1$)-bit Toffoli gate}
	To form an approximate Toffoli gate with $n$ control qubits, we choose the driving frequency $\Delta_0$ to be such that the zeroth qubit flips if and only if all control qubits are in the state $\ket{1}$, i.e., $\Delta_0 = \Delta_{11\ldots 1}$. \Cref{eq:Uxnot,eq:UxID} suggest that we have indeed obtained the aimed operation. 
	However, moving back from the rotating frame to the laboratory frame using $\hat{U}_{\vec{x},\text{lab}} = \hat{U}_\text{int}^\dagger(t) \hat{ U }_{\vec{x}}$ [see \cref{eq:Uint}], we encounter two discrepancies:
	(i) the additional phases $\exp(-i E_{x_0,\vec{x}} T)$ accumulated on each computational basis state due to $\hat{H}_\text{Ising}$, and
	(ii) the additional phase $-i$ in the resonant subspace $\vec{x}=1\ldots 1$. (note that this is \emph{not} a global phase).
	Note that such phases in the laboratory frame become relevant when subsequent non-commuting operations are performed. 
	
	The $2^{n+1}$ different energies $E_{x_0,\vec{x}}$ can in general be hard to compute for a large system. Undoing them may be even harder. However, one can conceive various specific configurations where resetting the phases is possible. In particular, whenever the Ising couplings $J_{jk}$ are symmetric under permutations on the control qubits, then the evolution depends only on the Hamming weight (the number of qubits in state $\ket{1}$) of the control qubits, which we define as $q = |\vec{x}|_H$. In such cases, only $n+1$ subspaces are unique, and hence only $n+1$ relative phases have to be considered. Various techniques can then be used to undo these dynamical phases. One example is to choose a total gate time $T$ such that all phases $E_{x_0,\vec{x}} T$ become a multiple of $2\pi$. For example, when all $J_{jk}$ are integer multiples of some energy scale $J$, then the values of $E_{x_0,\vec{x}}$ are also integer multiples of $2J$ such that a total driving time $T = 2 k \pi / J \ (k \in \mathbb{N})$ gets rid of unwanted phases. 
	Note that random experimental imperfections in $J_{jk}$ may still cause the fidelity of such phase recurrences to be affected. 
	A different strategy is to invert the sign of all $J_{jk}$ halfway through the protocol \cite{Groenland2019}. This undoes the accumulated phases, although care has to be taken to also change the phase $\theta$ of the resonant driving fields such that the previously caused rotation on the zeroth qubit is not counteracted. 
	


\begin{figure}
		\centering
		\includegraphics[width=.7\columnwidth]{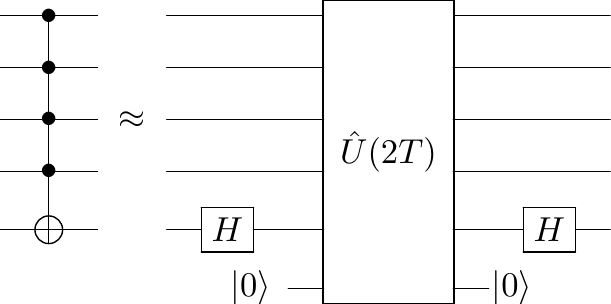}
		\caption{Circuit that turns two applications of the $i$-Toffoli (here indicated as the result of our protocol, with total time $2T = \pi / \Omega$ and arbitrary $\theta$) into a conventional Toffoli gate, at the cost of a single ancilla.}
		\label{fig:circuit}
	\end{figure}

	Assuming that we removed the phases due to $\hat{H}_\text{Ising}$, e.g. by transforming into the frame rotating with $\hat{H}_\text{Ising}$, we turn to removing the phase $-i$. This phase is a result of evolution by a Hamiltonian with trace $0$, which generates unitaries with determinant $1$. We will refer to the operation that acts as $-i\sigma^x$ on the target if and only if all controls are in the state $\ket{1}$ as the $i$-Toffoli. To turn this into a conventional Toffoli gate, we propose the circuit in \cref{fig:circuit}. Note that applying the resonant operation twice leads to a phase $-1$ in the resonant subspace. This is similar to a multiple-controlled $\sigma^z$-gate except that the sign is applied both when the target is in state $\ket{0}$ and when it is in state $\ket{1}$. Hence, we obtain a multiple-controlled-$\sigma^z$ gate which applies a sign $-1$ to the control qubits if and only if all these qubits are in the state $\ket{1}$. The state of the target is unimportant, and we may just as well initialize it to $\ket{0}$ before the protocol. Finally, the controlled-$\sigma^z$ gate is mapped to a controlled-$\sigma^x$ gate by using two Hadamard gates; these can be applied to any control qubit, which then takes the role of target of the resultant $(n-1)$-bit Toffoli gate.
	
	\subsection{Gate error and asymptotic scaling}\label{sec:gateError}
	Having dealt with the additional phases, we turn back to \cref{eq:UxID}, which we claim is a good approximation to the actual evolution in \cref{eq:timeEvo}. Surprisingly, the approximation error can be expressed using analytical methods. To achieve this, we assume a permutation symmetry between the control qubits, such that all couplings to the zeroth qubit are identical: $J_{0,k} = J$. In this case, the properties of each conserved subspace depend only on the Hamming weight $q$ of the controls. Using subscripts to denote the subspace in which the result is valid, we find that 
	\begin{align}
		\Delta_q = J (n-2 q).
	\end{align}
	Choosing the subspace with weight $q_0=n$ to be on-resonance, the detuning per subspace becomes $\delta = J (n-q)$. Finally, we set $T = \pi/2\Omega$ such that the resonant subspace is always completely inverted.
	
\begin{figure}
		\centering
		\includegraphics[width=\columnwidth]{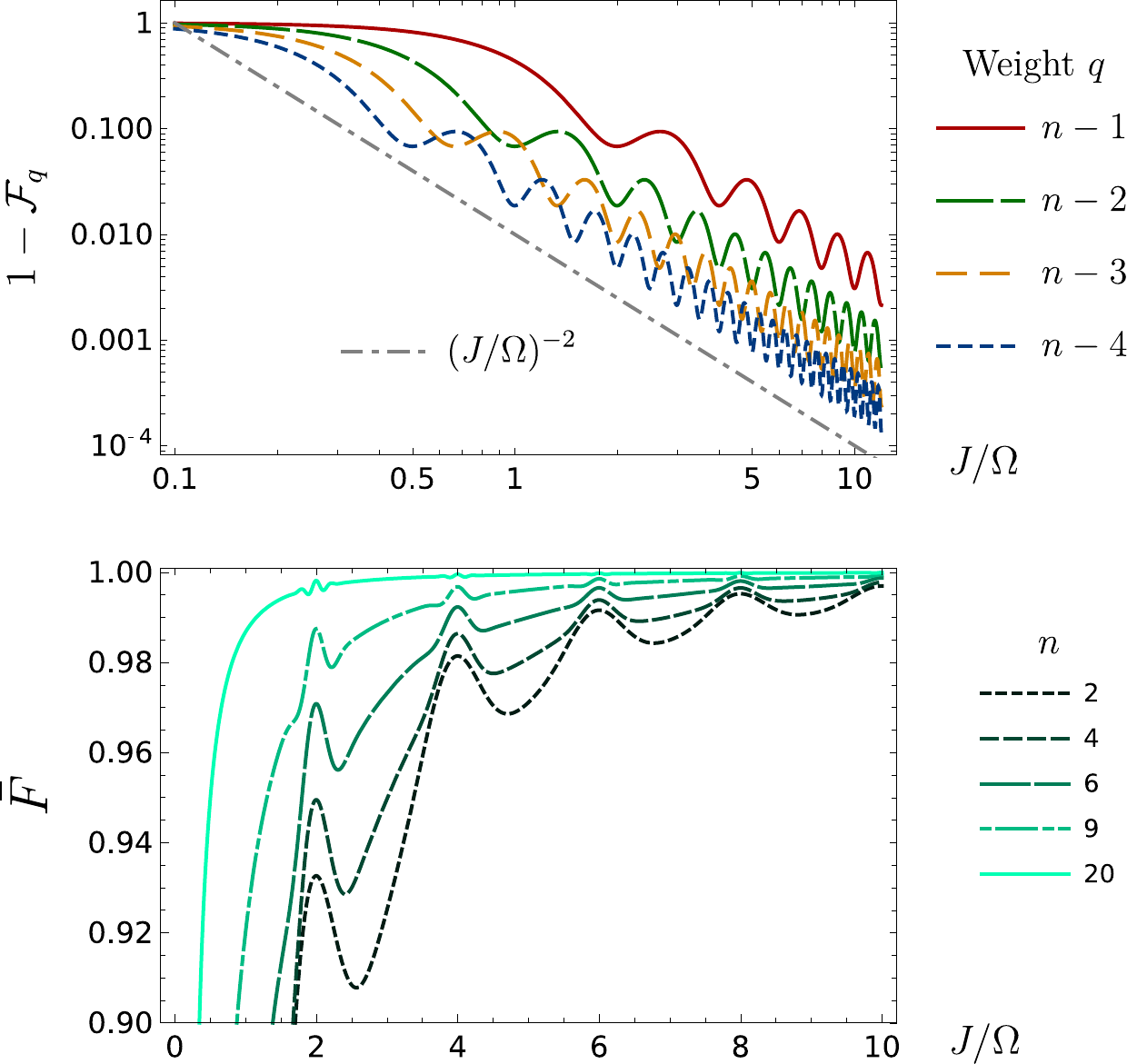}
		\caption{(a) the inner product error for subspaces with various Hamming weights. The error is seen to decay quadratically in $J/\Omega$. Note that the plot uses a log-log scale. (b) Process fidelity $\bar{F}$ for various values of $J / \Omega$ and for various system sizes. Both plots assume a permutation symmetry between control qubits and that the resonant subspace has weight $q_0 = n$.}
		\label{fig:Etrace}
	\end{figure}

	We can now calculate the accuracy of the approximation in \cref{eq:UxID} compared to \cref{eq:timeEvo} as a function of the system parameters. As metric, we consider the trace fidelity or matrix inner product per subspace:
	\begin{align*}
		\mathcal{F}_{q} (U_q, U_{\text{goal},q}) = & ~ \frac{1}{\text{dim}(U_q)} | \text{tr}( U_q U_{\text{goal},q}^\dagger ) | \\
		= & ~ \cos\left( \frac{\pi \gamma}{2} \right) \cos\left(\frac{\pi}{2} \sqrt{1+\gamma^2} \right)  \\
		& ~+ \frac{\gamma}{\sqrt{1+\gamma^2} } \sin\left( \frac{\pi \gamma}{2} \right) \sin\left( \frac{\pi}{2} \sqrt{1+\gamma^2} \right), 
	\end{align*}
	with $\gamma =  J (n-q)/\Omega$.
	This result is plotted \cref{fig:Etrace}(a) and is valid for arbitrary $n\geq 1$. Clearly, the subspace with weight $q=n-1$ is closest to resonance and therefore experiences the largest error. In each subspace, the fidelity scales approximately as $1-J^2/\Omega^2$, indicating that a sufficiently small $\Omega$ (hence larger gate time $T$) can, in theory, result in an $i$-Toffoli of arbitrary precision. 
	
	A metric for the overall gate fidelity uses a weighted sum over all subspaces, 
	\begin{align*}
		\mathcal{F}_\text{tr} (U, U_\text{goal}) &= \frac{1}{\text{dim}(U)}  \sum_{q=0}^n 2 \mathcal{F}_q \binom{n}{q}. 
	\end{align*}
	For comparison with later results that involve decoherence, we also introduce the process fidelity \cite{Nielsen2010,Nielsen2002,Horodecki1999,Schumacher1996}
	\begin{align}
		\bar{F} = &~ \int d\psi\langle \psi | \hat{U}^\dagger_\text{goal} \mathcal{C}(\psi)\hat{U}_\text{goal} | \psi\rangle,
		\label{eq:av_fidelity_formula}
	\end{align}
	where the integration is performed over all possible initial states $\ket{\psi}$ and $\mathcal{C}$ is the quantum channel that implements our protocol and outputs the resulting density matrix. In the following, we use this process fidelity as our metric of gate fidelity. In the special case that $\mathcal{C}$ is a unitary map, we can recycle our previously found trace fidelity \cite{Nielsen2002} using
	\begin{align*}
		\bar{F}   =    &~ \frac{  \text{dim}(U) \mathcal{F}_{\text{tr}}^2 + 1 }{ \text{dim}(U) + 1  }.
	\end{align*}
	The theoretical process fidelity of our driven $i$-Toffoli gate is plotted in the bottom panel of \cref{fig:Etrace} for varying $n$, where surprisingly the fidelity improves with larger system sizes. We explain this as follows. For any $n$, there are $n$ subspaces that differ in Hamming weight by $1$ from the resonant subspace with $q=n$, which are the least off-resonant. On the other hand, an exponentially large number of subspaces have a much larger off-resonance. Hence, the averaged error benefits more from the many off-resonant subsystems when $n$ increases.
	
	Note that, in the above, we worked in the interaction picture (\cref{eq:Uint}), such that the energies $E_{x_0, \vec{x}}$ dropped out. This allowed us to focus purely on the driving error as a function of $J/\Omega$. In fact, this calculation allows any couplings $J_{jk}$ between the control qubits, as these are not relevant in the interaction picture. Still, to make the $i$-Toffoli relevant to the lab frame, any relative phases need to be canceled somehow.
	
	The same derivation could be done for different definitions of gate fidelity or error, such as the operator norm error we discuss in \cref{app:opnorm}. This error considers only the worst possible input state to our protocol, and indeed we find that the error is independent of $n$ because it is fully determined by the subspace with weight $q=n-1$. 
	
	Finally, we note that one might consider the case where a subspace different from $q_0=n$ is to be flipped. By choosing the driving frequency $\Delta= J (n-2q_0)$, one will approximately find the operation where the zeroth qubit is rotated if and only if the control qubits have weight $|\vec{x}|_H = q_0$. The values $n-q$ in the above results should then be replaced with $q_0 - q$. 
	
	The asymptotic scaling of our protocol is surprising: the time required to perform our operation is independent of the number of qubits, and the induced error is either constant or (in the case of the trace fidelity) actually increases with a larger system size. This is a great improvement over conventional decompositions into one- and two-qubit gates, which take at least $O(\log(n))$ time and $O(n)$ gates. 
	
	Still, a critical look should be taken at our assumptions. First, strongly coupling $n$ qubits to a single target qubit would be challenging to realize in physical three-dimensional space, as each of these $n$ qubits would need to be sufficiently close to the zeroth.  Any realistic method to implement this would probably set a maximum for $n$. Second, the oscillation frequency $\sim \Delta_0$ for the resonant subspace $q_0 = n$ increases linearly with $n$ in our protocol. This may not be realistic, and if one requires the frequency $\Delta_0$ to be bounded, then the Hamiltonian energy scale needs to be scaled down by a factor $1/n$. This effectively causes all time scales to increase by a factor $n$, retrieving an $O(n)$ time protocol. Third, as $n$ increases so does the number of requirements. Evidently, $|\Delta_0 - \Delta_{\vec{x}}|\gg \Omega$ is more difficult to fulfill for all subspaces, except $\vec{x}'$, as the number of subspaces increases with $n$.
	
	Thus it is clear that there is a limit on how large $n$ can be. However, it is diffcult to predict how large as it will depend on the given implementation.

	\subsection{Simulation of the two-bit $i$-Toffoli gate with decoherence}\label{sec:example}
	
	In order to illuminate the performance of the system in a practical setting, we simulate our protocol for the $i$-Toffoli gate under realistic decoherence for $n=2$. We simulate the system using the Lindblad master equation and the interaction Hamiltonian of \cref{eq:HI} using the \textsc{qutip} \textsc{python} toolbox \cite{qutip}. The result is then transformed into the frame rotating with the diagonal of the Hamiltonian, and then the average process fidelity is calculated.
	
	\begin{figure}
		\centering
		\includegraphics[width=\columnwidth]{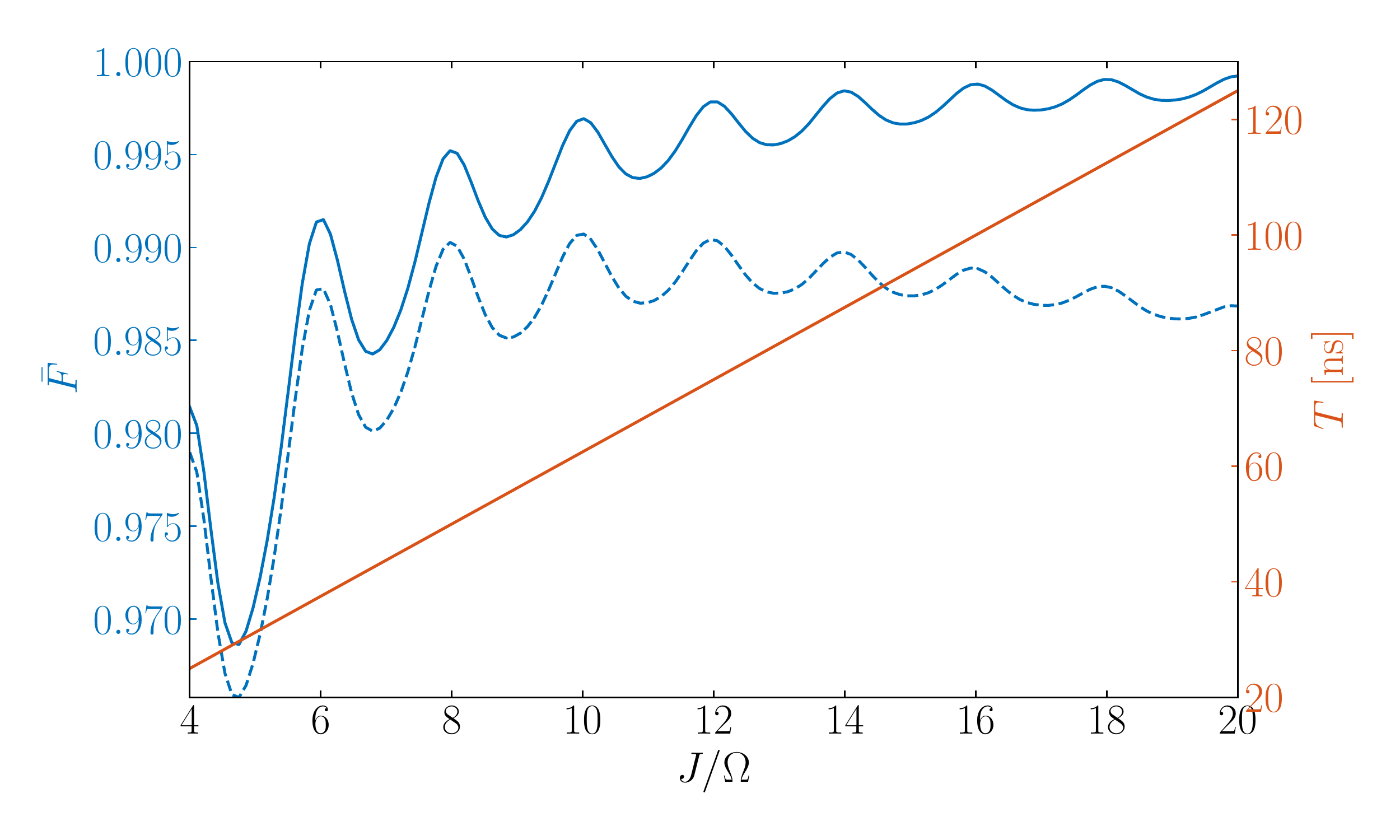}
		\caption{Simulation of the two-bit $i$-Toffoli gate for different values of the driving $J$. The straight red line indicates the gate time $T$ on the right $y$-axis, while the blue lines indicate the average fidelity on the left $y$ axis. The dashed blue line is the average fidelity with a decoherence time of $T_1 = T_2 =\SI{30}{\micro\s}$, while the solid line is without decoherence.}
		\label{fig:simulation}
	\end{figure}
	
	For all simulations we choose parameters which lie in a realistic range for a superconducting circuit experiment. However, our simulation is done for the general Hamiltonian, and is thus valid for any implementation with the same parameters. In particular we have $J_{0k}/2\pi = J/2\pi = \SI{40}{\MHz}$ and all other couplings are zero, while we change the Rabi frequency $\Omega/2\pi$ from 2 to \SI{10}{\MHz}. The average fidelity of the simulation can be seen in \cref{fig:simulation} together with the gate time. The figure shows the average fidelity both without any decoherence and with decoherence times of $T_1 = T_2 =\SI{30}{\micro\s}$ \cite{Wendin2017}, where $T_1$ indicates the relaxation time and $T_2$ indicates the dephasing time. Without any decoherence we find that the average fidelity increases asymptotically towards 1 as the driving decreases, with the only expense being an increase in gate time. Since decoherence increases over time, a longer gate time means lower fidelity, which is exactly what we observe when including decoherence in the simulations. In this case we find that the fidelity peaks just above $0.99$ at $J/\Omega = 8$, which yields a gate time of $T= \SI{62.5}{\nano\s}$. This fidelity is higher than any previously measured Toffoli gate fidelities \cite{Lanyon2008,Beterov2018}. However, we note that the fidelity is dependent on the parameters $J$ and $\Omega$, and thus changing these will change the fidelity.
	The oscillation of the average fidelity is due to a small mismatch in the phase of the evolved state compared to the desired gate, which disappears when $J/\Omega \in 2\mathbb{Z}$.
	
	As an indication of the fidelity of a conventional Toffoli gate, we simulate the same protocol for time $2T$ (see \cref{fig:circuit}), resulting in still above 0.98 fidelity. An additional two Hadamard gates should then still be applied, but we remain agnostic to the errors these would introduce.
	
	\begin{figure}
		\centering
		\includegraphics[width=\columnwidth]{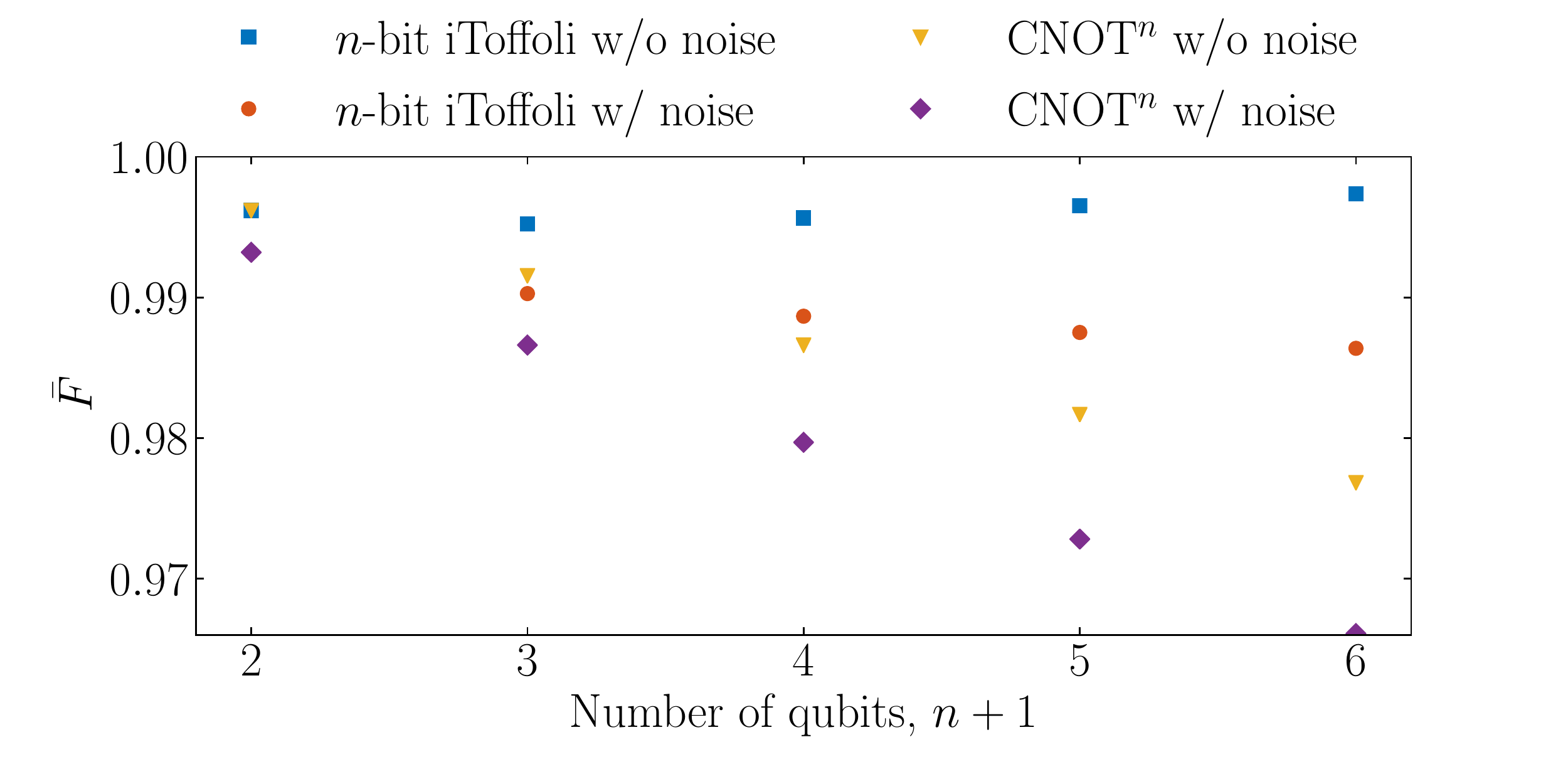}
		\caption{Average fidelity as a function of the number of qubits for the $n$-bit $i$-Toffoli gate and the \textsc{cnot}$^n$ gate. Simulations done with noise have a decoherence time of $T_1 = T_2 =\SI{30}{\micro\s}$. All simulations are done with $J/\Omega=8$, i.e., peak fidelity (cf. \cref{fig:simulation}). Note that the one-bit $i$-Toffoli and \textsc{cnot}$^1$ are the same gate, which is an example of a Barenco gate. 
		}
		\label{fig:control}
	\end{figure}
	
	We investigate the peak fidelity of the $n$-bit $i$-Toffoli gate as a function of the number of control qubits. This is done by simulating the gate for different $n$ but with $J/\Omega = 8$ in all cases. The result are shown in \cref{fig:control}. We find that when we do not include the decoherence of the qubits the average fidelity [found using \cref{eq:av_fidelity_formula}] increases when there are more than two control qubits and we stay above 0.995 in fidelity for all cases. This is in agreement with the analytical result discussed in \cref{sec:gateError}. When decoherence is included the fidelity decreases as the number of qubits increases as one would expect. Thus we conclude that the major contribution to error in the scheme is the decoherence of the qubits.

	\section{A single control, multiple inversion gate}\label{sec:cnotn}
	
	Multiple applications of a controlled-\textsc{not} gate on several different qubits, with the same qubit controlling all the gates, are essential in many aspects of quantum information, particularly in error correction such as Shor's code \cite{Nielsen2010}. We therefore present a scheme for implementing inverting multiple qubits with the same control qubit in a single step. We will refer to this scheme as a \textsc{cnot}$^n$-gate.
	
	Starting with $n+1$ qubits, we employ the same overall Hamiltonian as in \cref{sec:cnnot}, $\hat H = \hat H_0 + \hat H_\text{Ising} + \hat H_\text{drive}$, where $\hat H_0$ and $\hat H_\text{Ising}$ are given in \cref{eq:H0,eq:Ising}, respectively, while we require $J_{jk} = 0$ for $k>0$. The driving Hamiltonian is now given as
	\begin{equation}
	\hat H_\text{drive} = \sum_{j=1}^n\left[\alpha_j(t)\sigma^x_j + \beta_j(t)\sigma^y_j\right].
	\end{equation}
	where the driving fields are given as in \cref{eq:Bi}. This is essentially the same system as in \cref{sec:cnnot} but now with the driving on what was before called the control qubits. We therefore denote our quantum states in the same way as before, $|x_0,\vec x \rangle $; however, now we are interested in flipping the qubits in the state $\vec x$ conditional on the state of the zeroth qubit $x_0$.
	This means that the Hilbert space only decomposes into two conserved subspaces, one spanned by $\{|0,\vec x\rangle\}$, and one spanned by $\{|1,\vec x\rangle\}$. We now transform into a rotating frame using the transformation
	\begin{equation}
	\hat U_\text{int} (t) = \exp\left( i \left[ \hat H_0 + \frac{1}{2}\sum_{j=1}^n J_{j0} \sigma_j^z\sigma_0^z\right]t\right).
	\end{equation}
	In this frame the Hamiltonian takes the form
	\begin{equation}
	\begin{aligned}
	\hat H_I =& \sum_{j=1}^n \Omega\left\lbrace\sigma^x_j \cos\left[(\Delta_j - J_{j0}\sigma_0^z)t +\theta_j\right] \right. \\
	&+ \left.\sigma^y_j \sin \left[(\Delta_j - J_{j0}\sigma_0^z)t +\theta_j\right]\right\rbrace,
	\end{aligned}
	\end{equation}
	from which we see that we obtain selective inversion of the $n$ qubits, at time $T=(2m+1)\pi/2\Omega$, if we require $\Delta_j = -J_{j0}$.
	The time evolution operator takes the form
	\begin{align}
		\hat U(t=T) = & |0\rangle\langle 0|_0 \bigotimes_{j = 1}^{n} \hat{I}_{j} \\ & + (-i)^n|1\rangle\langle 1|_0  \bigotimes_{j = 1}^{n} \left( \sigma^x_{j}\cos\theta_j + \sigma^y_{j}\sin\theta_j\right),\nonumber
	\end{align}
	where $\hat{I}_{j}$ is the identity of the $j$th qubit and $|0\rangle\langle 0|_0$ and $|1\rangle\langle 1|_0$ operate only on the zeroth qubit. We find that the phase $(-i)^n$ on the inverting part of the operator is now dependent on the number of target qubits. In this case, it is easily canceled by a single-qubit phase gate of the form $\text{diag}(1, i^n)$ on the control qubit. Note that in the case of an even number of qubits the phase is either $\pm 1$, which can be taken care of by choosing the right phases $\theta_j$, in which case the single-qubit phase gate is unnecessary.
	
	Since the only difference between the Hamiltonian of the \textsc{cnot}$^n$ gate and the $n$-bit Toffoli gate in \cref{sec:cnnot}, is which qubits are being driven, a numerical simulation of the \textsc{cnot}$^n$ gate as a function of the ratio $J/\Omega$ yields an average fidelity comparable to the one for the $n$-bit $i$-Toffoli gate in \cref{fig:simulation}. However, the \textsc{cnot}$^n$ gate has a slightly lower fidelity since more qubits are now inverted. The peak average fidelity can be seen in \cref{fig:control}, where the average fidelity decreases as a function of the number of qubits. This behavior is expected since the \textsc{cnot}$^n$ gate does not approximate the identity better for larger $n$. Note that the one-bit $i$-Toffoli gate is the same as the \textsc{cnot} gate, which is why the average fidelities are identical in this case. This is also the fidelity one gets when simulating the Barenco gate in\cref{sec:Barenco}.

	\section{Experimental implementations}\label{sec:exerimental}
	
	The ideas presented here are applicable in various quantum information technologies. Our main focus is on superconducting circuits, which we discuss below and elaborate on in \cref{app:analCircuit}. We also discuss the prospects of implementing our operation using Rydberg atoms and trapped ions. 
	
	\subsection{Superconducting circuit implementation}\label{sec:SCI}
	
	An implementation of our $n$-bit Toffoli gate would require quite large longitudinal $ZZ$ couplings, in the sense that they must dominate over the transversal $XX$ couplings. For superconducting circuits this regime is within experimental reach according to Ref.~\cite{Kounalakis2018}.
	Inspired by the superconducting circuit realizing the coherent quantum router in Ref.~\cite{Christensen2019}, we propose to implement the $n$-bit Toffoli gate and the \textsc{cnot}$^n$ gate by connecting $n$ transmon qubits \cite{Koch2007,Schreier2008} via Josephson junctions (with as small a parasitic capacitance as possible) to another transmon qubit, which we call the zeroth transmon, in correspondence with naming of the qubits in \cref{sec:cnnot}. Such a circuit has the Hamiltonian
	\begin{equation}\label{eq:HCircuit}
	\begin{aligned}
	\hat H_\text{circ} =& \frac{1}{2} \vec{\hat p}^T K^{-1} \vec{\hat p} - \sum_{j=0}^{n} E_j\cos\hat\varphi_j \\
	&-  \sum_{j=1}^n E_z\cos(\hat\varphi_{0}-\hat\varphi_j) ,
	\end{aligned}
	\end{equation}
	where $\hat\varphi_j$ are the node fluxes and $\vec{\hat p}^T = (\hat p_0,\hat p_1,\dots,\hat p_n)$ are the conjugate momenta, fulfilling the commutator relation $[\hat \varphi_j,\hat p_k] = i\delta_{jk}$, with $\delta_{jk}$ is the Kronecker delta. In addition $K$ is the capacitance matrix of the circuit. Examples of circuits and capacitance matrices, for $n=2$ and 3, can be found in \cref{app:analCircuit}.
	The capacitive couplings, coming from the parasitic capacitances of the Josephson junctions, yield transversal $XX$ couplings between all the qubits in the model. We are however not interested in these couplings, and thus we require capacitances of the Josephson junctions to be much less than the transmon capacitances, which will leave the capacitance matrix being approximately diagonal, effectively suppressing undesired transversal $XX$ couplings between the control qubits stemming from the capacitances. We further detune the zeroth qubit from the control qubit such that the remaining $XX$ couplings are suppressed. This leaves only longitudinal $ZZ$ couplings as desired. When truncating the Hamiltonian in \cref{eq:HCircuit} to an Ising-type model one reaches the non driving term of the Hamiltonian in \cref{eq:HI}. We obtain the driving part of the Hamiltonian by applying a microwave field to the desired qubits, depending on whether we want to realize the $n$-bit Toffoli gate or the \textsc{cnot}$^n$ gate. A detailed calculation going from the circuit design to the gate Hamiltonian can be found in \cref{app:analCircuit}.
	
	\subsection{Rydberg atoms}
	Ultracold atoms of the Rydberg type natively feature a strong Ising-type interaction \cite{Saffman2016}, making these a promising platform for our protocol to be implemented. Various earlier proposals for multiqubit operations based on the Rydberg blockade interaction exist, such as Refs. \cite{Lukin2001,Unanyan2002,Isenhower2011,Shi2018}, and some of these have been experimentally tested \cite{Ebert2015,Zeiher2015}. The specific Toffoli-type gates have never been implemented. However, Refs. \cite{Gulliksen2015,Petrosyan2016} perform detailed simulations of previous proposals for driven protocols for Toffoli gates in the context of quantum algorithms, finding that the multiqubit implementation may have advantages over a sequence of one- and two-qubit gates. We hope that this motivates future work to consider our protocol on a Rydberg atom quantum computer in more detail.

	\subsection{Trapped ions}
	Trapped ions are very well suited to simulate the Ising model with all-to-all connectivity. In these systems, linear crystals of ions are held in electric traps, and for each ion, two electronic states are chosen to form the qubit degree of freedom. By coupling these qubits to motional states of the ions using lasers, an effective interaction between the qubits can be formed, which approximates the Ising model. The spin-spin couplings can be made of the form $J_{jk} \propto 1 / | j - k |^\alpha$ with $\alpha \in [0,3]$ \cite{Kim2009,Blatt2012,Britton2012,Islam2013}. The choice $\alpha=0$ makes all interactions equal, leading to a highly symmetric system for which the energies $\bar{E}_{\vec{x}}$ are efficiently calculated. 
	
	We identify various challenges for an implementation of our gate using trapped ions. 
	Firstly, the amplitude of the Ising interaction $J$ is determined by the coupling strength of the lasers to the motional excitation of the ions. However, to prevent qubit-motion entanglement, this laser coupling has to be rather weak, such that only virtual phonon excitation occurs~\cite{Kim2009}. Typically interaction strengths between the qubits lie in the kilohertz range in these systems. Since our resonant field on the ancilla must have an amplitude $\Omega$ that is even much smaller than $J$, this leads to very long gate times, well beyond a millisecond. To illustrate, a closely related experiment was performed in Ref. \cite{Senko2014}. Here, up to 18 ions are made to approximate an Ising interaction, while at the same time, a resonant driving field is applied to all ions simultaneously. The effective interaction strength $J$ is indeed of the order of a kilohertz, while the driving field amplitude is a fraction of that. While this is sufficiently strong to identify transitions for spectroscopic applications, the resulting evolution would likely be too slow to perform a high-fidelity multiqubit gate. 
	
	Alternatively, we can tune the laser frequencies closer to the eigenfrequencies of the ion motion, such that the phonon excitation is non dispersive~\cite{Kim2009}. In this situation, we would have to choose the gate time and driving field such that the phonon populations exactly return to their initial state at the end of the gate sequence. This ensures that the qubit states get disentangled from the ion motion. In this scheme, $J$ is greatly increased, and coupling strengths $J$ in excess of 100~kHz can be obtained~\cite{Schafer:2018}. However, the phonon numbers oscillate with large amplitude during this type of gate sequence. Driving a transition between the states $\ket{1,\vec{x}}$ and $\ket{0,\vec{x}}$ while these are entangled to different phonon states introduces errors in the final gate. In particular, disentangling the qubit and motional states when also applying the driving field $\Omega$ is not trivial. It is worth investigating whether newly developed techniques for finding robust gate operations using pulse engineering could be successfully applied to this problem~\cite{Palmero:2017,Manovitz:2017,Schafer:2018,Leung:2018,Webb:2018,Shapira:2018}. Stroboscopic techniques could also be used such that the driving field is only applied at times when the qubit gets disentangled from the motion~\cite{Lanyon:2011}. We conclude that the indirect nature of the trapped ion's Ising interaction introduces several obstacles that have to be bridged before our protocol could be competitive with conventional gate decompositions. 

	\section{Applications in quantum error correction}\label{sec:QEC}
	
	In this section we discuss how to use the results above to create an efficient error correction code. We consider the three-qubit bit-flip code \cite{Nielsen2010} and the Steane seven-qubit code \cite{Steane1996a,Steane1996b}. We focus on bit flip rather than phase errors in the three-qubit code, since the decay time for relaxation is usually half the decay time for dephasing in the case of transmons \cite{Schreier2008,Houck2009}. One can however easily change the code into correcting phase errors by applying Hadamard gates around the source of error \cite{Nielsen2010}. This could be useful in an implementation of a 0-$\pi$ qubit, which has a long relaxation time but a rather short dephasing time \cite{Kitaev2006,Brooks2013,Gyenis2019}. 
	The three-qubit code has previously been implemented using superconducting circuits to a fidelity of 0.85 \cite{Reed2012} and with trapped ions to a fidelity of approximately 0.98 \cite{Chiaverini2004}.
	The Steane seven-qubit code has been implemented with a state fidelity between 0.85 and 0.95 using trapped ions \cite{Nigg2014}.
	
	In the following all simulations are done without worrying about the phase generated by the inverting, i.e., it is done with the $i$-Toffoli gate, as it is irrelevant for the encoding.
	The error correction codes is simulated using the Lindblad master equation using the \textsc{qutip} \textsc{python} toolbox \cite{qutip}. All Ising couplings are assumed to be $J/2\pi = \SI{40}{\MHz}$.

	\subsection{Three-qubit bit flip code}
	
\begin{figure}
		\begin{equation*} \Qcircuit @C=2 em @R=1.4 em @!R {
				\lstick{1\quad\,\ket{0}}  & \targ & \multigate{2}{\rotatebox{90}{ERROR}} & \targ & \ctrl{1} & \qw\\
				\lstick{2\quad\, \ket{0}}  &  \targ & \ghost{E} & \targ & \ctrl{1} & \qw\\
				\lstick{3\quad \ket{\psi}} & \ctrl{-2} & \ghost{E} & \ctrl{-2} & \targ & \qw} \end{equation*}
		\caption{Effective three-qubit error correction code using two \textsc{cnot}$^2$ gates and a Toffoli gate. We label the top qubit 1, the middle qubit 2, and the lowest qubit 3. Note that the figure is shown with regular Toffoli gates, while our simulation is done with the $i$-Toffoli gates; however, it does not change the result.}
		\label{fig:QEC}
	\end{figure}
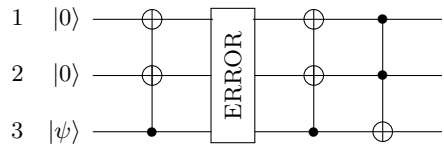

	The original three-qubit bit flip code works by first applying two \textsc{cnot} gates before the error source, and then two \textsc{cnot} gates after the error followed by a single two-bit Toffoli gate. This means a total of five steps. However, using our results the code can be performed in merely three steps: apply a single \textsc{cnot}$^2$ gate before the source of error, a single \textsc{cnot}$^2$ gate after the error, and finally a single $n$-bit Toffoli gate. A quantum circuit of the error correcting code can be seen in \cref{fig:QEC}.
	
\begin{figure}
		\centering
		\includegraphics[width=\columnwidth]{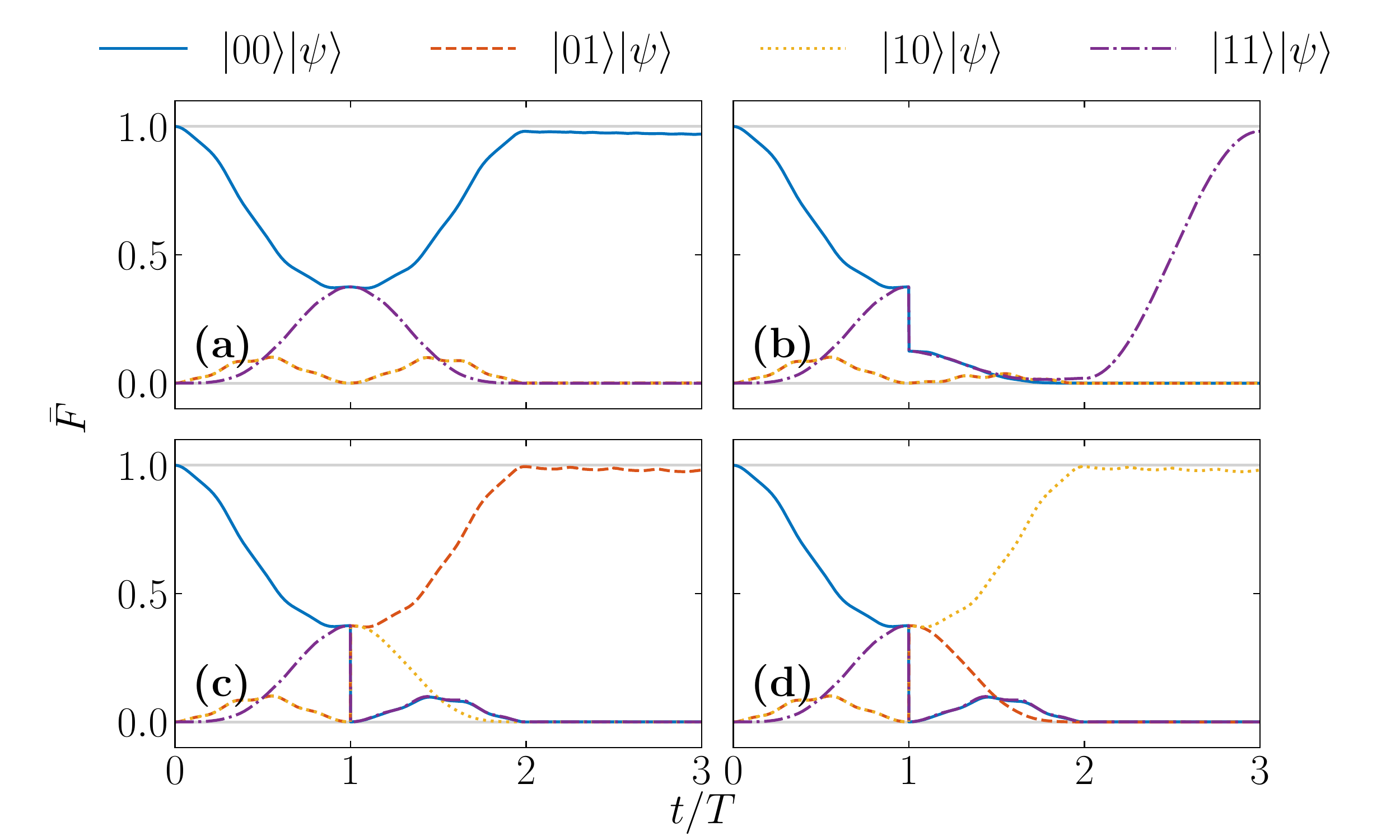}
		\caption{Average fidelity [\cref{eq:av_fidelity_formula}] of different states found by simulating the quantum error correction code shown in \cref{fig:QEC} using the gates developed in the previous sections: (a) no error, (b) error on the first qubit, (c) error on the second qubit, (d) error on the third qubit.}
		\label{fig:QECsimulation}
	\end{figure}

	The first two qubits are initiated in the state $\ket 0$, while the third qubit is initiated in the normalized state
	\begin{equation}\label{eq:psi}
	\ket \psi = \alpha \ket 0 + \beta \ket 1.
	\end{equation}
	The system is then operated as a \textsc{cnot}$^2$ gate by driving the first two qubits with an Rabi frequency of $\Omega = J/8$ for one period, i.e., $T = \SI{50}{\nano\s}$. After this a bit-flip error might occur. This is followed by another driving of the two first qubits for one period. Finally the last qubit is driven for one period. All this is done in $\SI{150}{\nano\s}$. By averaging over the Bloch sphere for the input state $\ket \psi $ in \cref{eq:psi} we find the average fidelity of the code. In \cref{fig:QECsimulation} we present the average fidelities for the three-qubit error correction code for a single bit flip on the different bits. From the simulation we see that the error is corrected with a fidelity above 0.99.
	
	\subsection{Steane code}
	
\begin{figure}
		\begin{equation*} \Qcircuit @C=1.5em @R=1em @!R  {
				& \text{(a)} & \text{(b)} & \text{(c)} & \text{(d)} & \\
				\lstick{1\quad\ket{+}}  & \qw & \qw & \qw & \ctrl{6} & \qw\\
				\lstick{2\quad\ket{+}}  & \qw & \qw & \ctrl{5} & \qw & \qw\\
				\lstick{3\quad\ket{+}}  & \qw &  \ctrl{4} & \qw & \qw & \qw \\
				\lstick{Q\quad \ket{\psi}}  & \ctrl{2} & \targ & \targ & \qw & \qw\\
				\lstick{4\quad\, \ket{0}}  & \targ & \targ & \qw & \targ & \qw\\
				\lstick{5\quad\, \ket{0}}  & \targ & \qw & \targ & \targ& \qw \\
				\lstick{6\quad\, \ket{0}}  & \qw & \targ & \targ & \targ& \qw } \end{equation*}
		\caption{Encoding scheme for the Steane code. Qubit $Q$ is initially in state $\ket{\psi}$, and the circuit encodes it into a seven-qubit state using one \textsc{cnot}$^2$ gate and three \textsc{cnot}$^3$ gates. The first three qubits are prepared in the state $|+\rangle = (|0\rangle + |1 \rangle )/\sqrt{2}$, while the last three qubits are prepared in the state $|0 \rangle$.}
		\label{fig:steane}
	\end{figure}
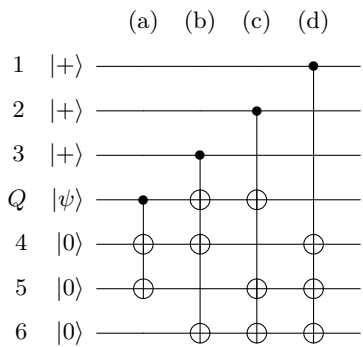

	The Steane code is a bit more intricate than the three-qubit code as it involves encoding on seven qubits. This is two more than the minimum number of qubits needed for protection against both bit-flip and phase errors \cite{Nielsen2010}, but it is the simplest Calderbank-Shor-Steane (CSS) code (stabilizers built from only either $Z$ or $X$ rotations) which protects against both bit-flip and phase errors. The encoding scheme for the Steane code can be seen in \cref{fig:steane}.
	
\begin{figure}
		\centering
		\includegraphics[scale=0.3]{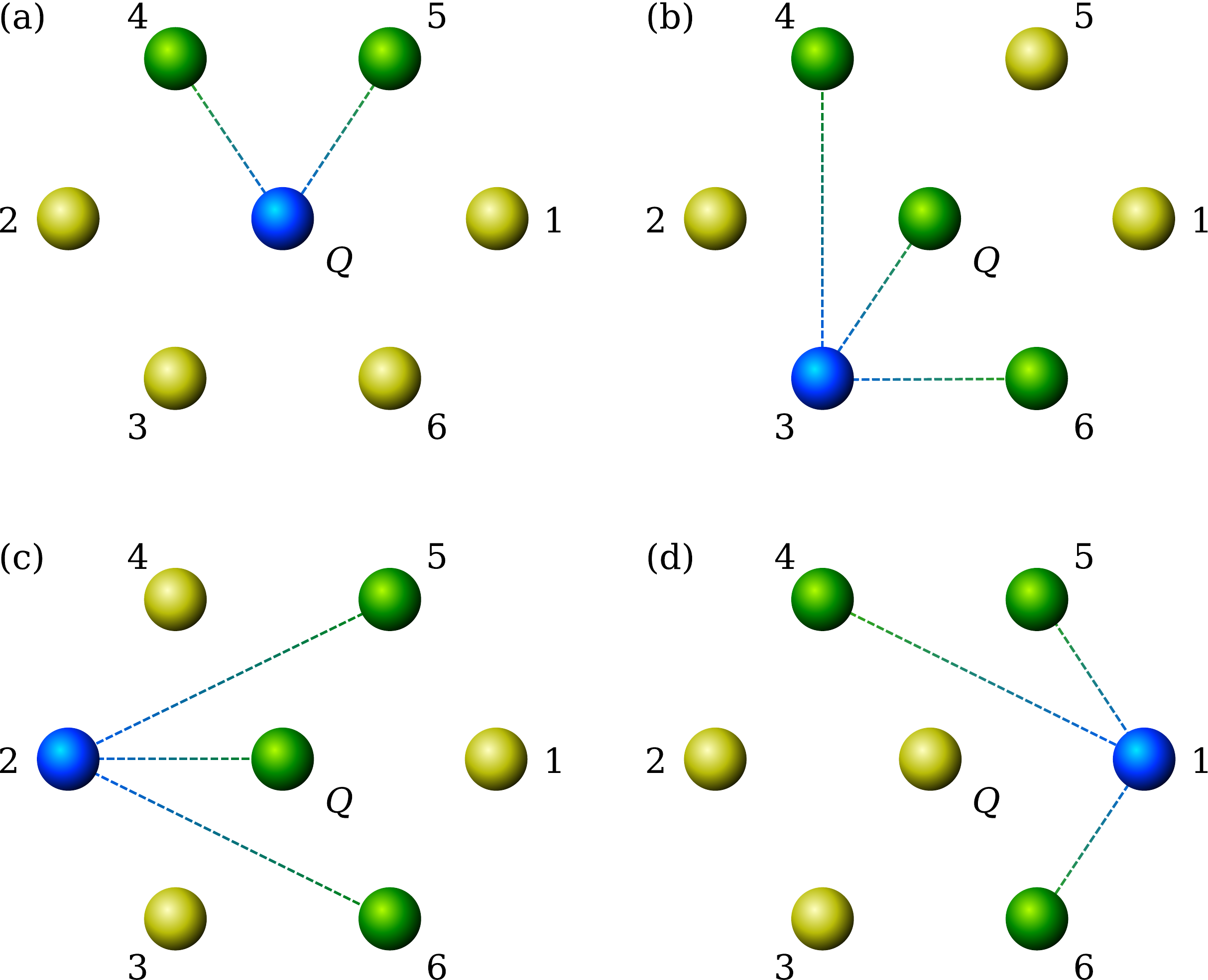}
		\caption{The four steps realizing the Steane code using \textsc{cnot}$^n$ gates. Green spheres (spheres with one connection) represent target qubits, i.e., qubits on which the \textsc{not}-operation is performed, blue spheres (spheres with multiple connections) are control qubits, and the yellow spheres (unconnected spheres) represents idle qubits. The four steps (a)-(d) corresponds to the four gates in \cref{fig:steane}.}
		\label{fig:SteaneStep}
	\end{figure}

	As the encoding scheme only uses \textsc{cnot}$^2$ and \textsc{cnot}$^3$ gates, it is necessary to be able to perform gate operations on some of the seven qubit but not all. This can be achieved \emph{in situ} in superconducting circuits by varying the magnetic flux through the Josephson junctions which connects qubits which are desired unconnected. An overview on how to connect the seven qubits in the four steps of the encoding can be seen in \cref{fig:SteaneStep}. Using the regular \textsc{cnot} gate, the Steane encoding takes 11 steps, while with \textsc{cnot}$^n$ gates it can be done in just four steps.
	
	Seven qubits are initialized, three in the state $|+\rangle = (|0\rangle + |1 \rangle )/\sqrt{2}$ and three in the state $|0 \rangle$, while the last qubit is prepared in the state of \cref{eq:psi}. The driving of the target qubits is the same for all steps, yielding a total time of $4T$ for the encoding. We average over the Bloch sphere for the input state $|\psi\rangle$ in order to find the average fidelity. The fidelity is found by taking the overlap between the seven-qubit output state and the state $\alpha |0\rangle_L + \beta |1\rangle_L$, where the expressions for the two states $|0\rangle_L$ and $|1\rangle_L$ are the appropriate encoding states for the Steane code, when the encoding is done with $i$-Toffoli gates. We absorb the additional phases $i^n$ that come with our driven protocol into the definitions of $|0\rangle_L$ and $|1\rangle_L$, as defined in \cref{app:steane}. These additional phases do not change the error correcting properties of the code. 
	
\begin{figure}
		\centering
		\includegraphics[width=\columnwidth]{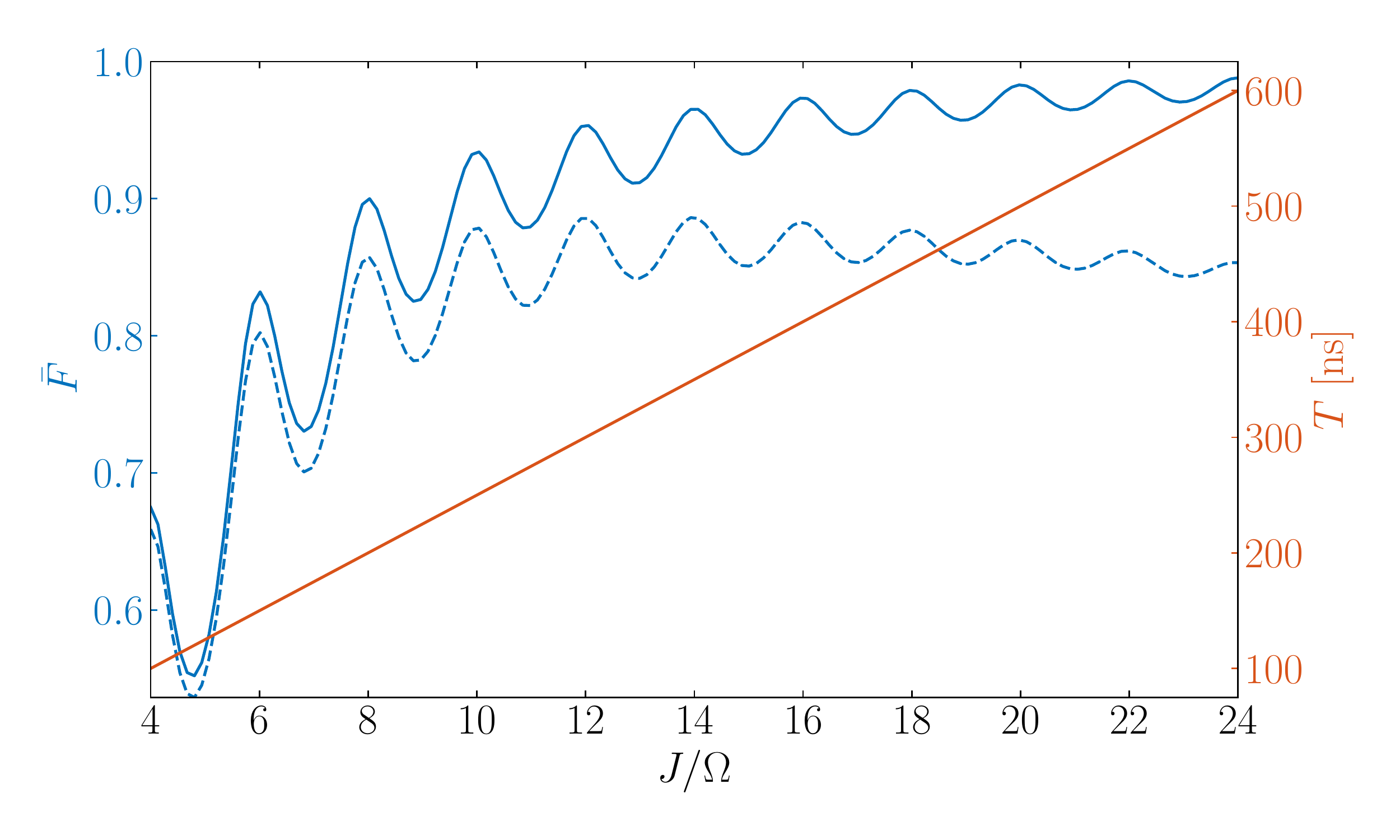}
		\caption{Simulation of the Steane encoding scheme seen in \cref{fig:steane}, using \textsc{cnot}$^n$ gates. The simulation is done for $J/2\pi = \SI{40}{\MHz}$. The straight red line indicates the gate time $T$ on the right $y$-axis, while the blue lines indicate the average fidelity on the left $y$ axis. The dashed blue line is the average fidelity with a decoherence time of $T_1 = T_2 =\SI{30}{\micro\s}$, while the solid line is without decoherence.}
		\label{fig:steaneSimu}
	\end{figure}

	The results of the simulation can be seen in \cref{fig:steaneSimu}. The result is similar to the one presented in \cref{fig:simulation}, however, with longer gate times and lower fidelities. When not considering decoherence, the lower fidelity is also a result of the fact that we need four gates, and thus the infidelities of all gates accumulate. The fidelity peaks just below 0.9 when including decoherence in the simulation, which is a lower number than before because more qubits are subjected to decoherence. The longer gate time is a result of the fact that we are now dealing with four gates, compared to one in \cref{fig:simulation}. However, this is still a rather short time compared to if we had only used two-qubit gates, which would increase the gate time by almost a factor of 3.
	
	\section{Conclusion and outlook}\label{sec:conclusion}
	
	We proposed a simple single-step implementation of $n$-bit Toffoli gates, \textsc{cnot}$^n$-gates, and the Barenco gate and showed that these exhibit a high fidelity, with the main cause of error being the qubits' decoherence. These gates can easily be transformed into C$^n$Z or CZ$^n$ gates by applying Hadamard gates on the target qubits. While the difficulty of implementing our gates does increase with $n$, we believe that our gates can provide many advantages to certain types of quantum computers, especially compared to rather deep equivalent circuits built from one- and two-qubit gates. 
	As an example of an implementation of the gates for quantum information processing we discussed superconducting circuit design of the gates and possible implementation in Rydberg atoms and trapped ions, though the idea is not limited to these quantum information schemes. 
	By simulating the protocol we showed that the gates can easily be concatenated into error correction codes. The gates proposed in this paper are not limited to the three-qubit error correcting code or the Steane code. They can be applied to numerous other codes making them more effective.
	These results could enhance the performance of near-term quantum computing experiments on algorithms that require many Toffoli gates or same-control \textsc{cnot} gates.

	\begin{acknowledgements}
		The authors would like to thank K. S. Christensen, T. Bækkegaard, L. B. Kristensen, and N. J. S. Loft for discussion on different aspects of the work.
		S.E.R and N.T.Z were supported by the Danish Council for Independent Research and the Carlsberg Foundation.
		K.G and K.S were supported by the QM\& QI grant of the University of Amsterdam, supporting QuSoft. 
		R.G was supported by the Netherlands Organization for Scientific Research (Grant No. 680.91.120).
	\end{acknowledgements}

	\onecolumngrid
	\appendix
	
	\section{Analysis of the superconducting circuit}\label{app:analCircuit}
	
\begin{figure}
		\centering
		\includegraphics[scale=0.45]{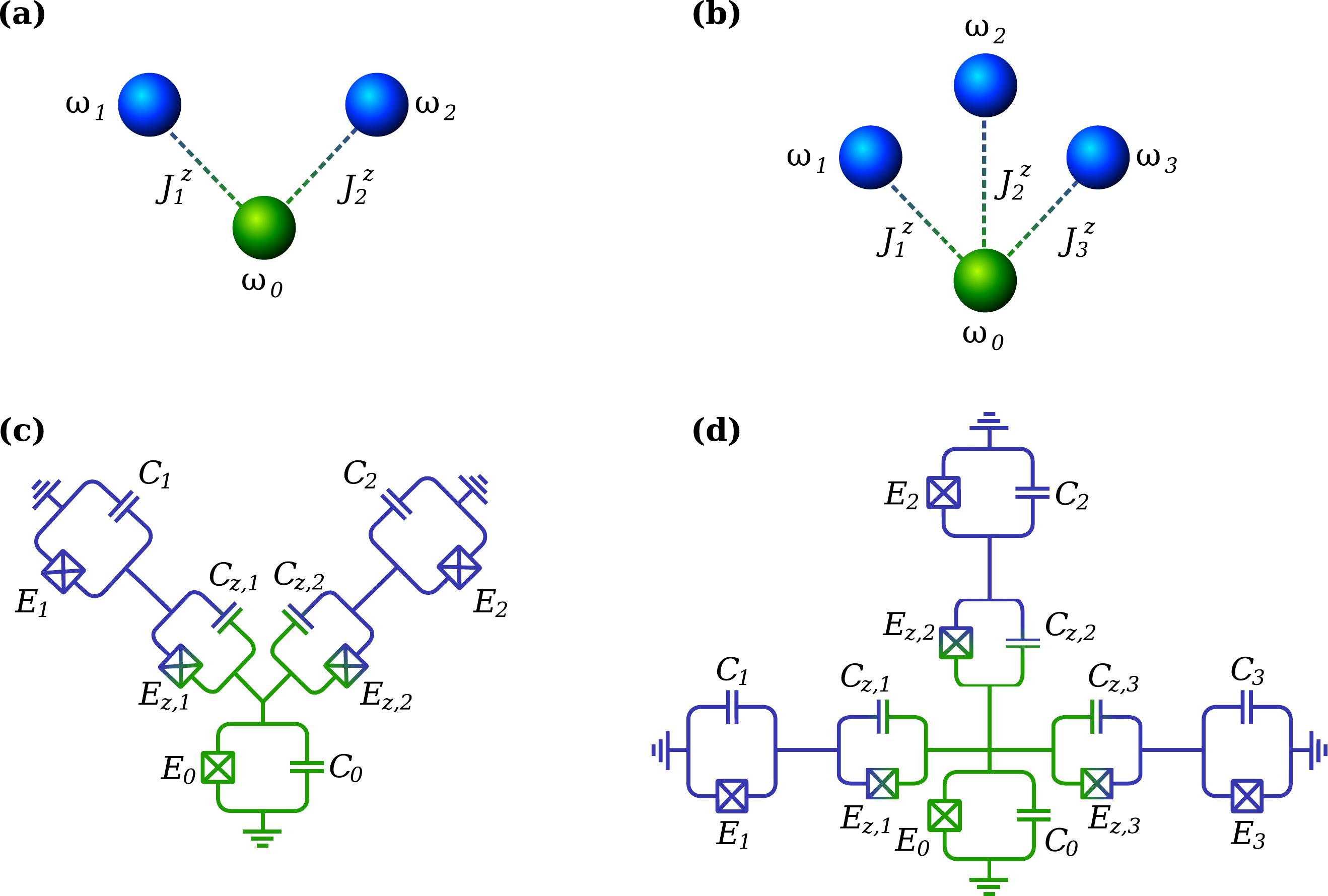}
		\caption{Implementation of the (a) two-bit and (b) three-bit $i$-Toffoli gates in superconducting circuits with the green spheres ($\omega_0$) representing the target qubit and the blue spheres representing the control qubits. Also shown are (c) and (d) the superconducting circuits yielding the models in (a) and (b), respectively. The different parts of the system are colored according to their role, as per (a) and (b).}
		\label{fig:CircuitDiagram}
	\end{figure}

	Following the procedure of Refs. \cite{Devoret1997,Devoret2017} we obtain the Lagrangian from the circuit diagrams in \cref{fig:CircuitDiagram}
	\begin{equation}
	L = 2 \sum_{i=0}^{n} C_i\dot\varphi_i^2 + 2 \sum_{i=1}^{n} C_{z,i}\left(\dot\varphi_i - \dot \varphi_{0} \right)^2 +  \sum_{i=0}^{n} E_i\cos\varphi_i +  \sum_{i=1}^n E_{z,i}\cos(\varphi_{0}-\varphi_i),
	\end{equation}
	where $\hat{\varphi}_i$ are the node fluxes across the Josephson junctions of the respective qubits. The first two terms come from the capacitors and are interpreted as kinetic terms, and the remaining terms come from the Josephson junctions and are interpreted as potential terms. The $n$ indicates the number of blue transmon qubits on the circuit diagram, i.e., in \cref{fig:CircuitDiagram}(c) $n=2$.
	
	The Lagrangian can be rewritten into the Hamiltonian in \cref{eq:HCircuit} doing the usual Legendre transformation. The capacitance matrix in the two-bit case is then [\cref{fig:CircuitDiagram}(c)]
	\begin{equation}
	K = \begin{bmatrix}
	C_0 + C_{z,1} + C_{z,2} & -C_{z,1} & -C_{z,2} \\
	-C_{z,1} & C_1 + C_{z,1} & 0 \\
	-C_{z,2} & 0 & C_2 + C_{z,2}
	\end{bmatrix}.
	\end{equation}
	while in the three-bit case [see \cref{fig:CircuitDiagram}(d) for a circuit diagram of the three-bit case] it becomes
	\begin{equation}
	K = \begin{bmatrix}
	C_0 + C_{z,1} + C_{z,2} + C_{z,3} & -C_{z,1} & -C_{z,2} & -C_{z,3} \\
	-C_{z,1} & C_1 + C_{z,1} & 0 & 0 \\
	-C_{z,2} & 0 & C_2 + C_{z,2} & 0 \\
	-C_{z,3} & 0 & 0 & C_3 + C_{z,3}
	\end{bmatrix},
	\end{equation}
	and so on for higher $n$. The typical transmon has a charging energy much smaller than the junction energy, and therefore the phase is well localized near the bottom of the potential. We can therefore expand the potential part of the Hamiltonian to fourth order
	\begin{align*}
		U(\varphi) =& \sum_{i=0}^{n} E_i \left[\frac{1}{2}\varphi_i^2 - \frac{1}{24}\varphi^4_i\right] + \sum_{i=1}^{n} E_{z,i} \left[\frac{1}{2}\left(\varphi_i - \varphi_{0}\right)^2 - \frac{1}{24}\left(\varphi_i - \varphi_{0}\right)^4\right] \\
		=& \sum_{i=0}^{n} E_i \left[\frac{1}{2}\varphi_i^2 - \frac{1}{24}\varphi^4_i\right] + \sum_{i=1}^{n} E_{z,i} \left[\frac{1}{2}\left(\varphi_i^2 + \varphi_{0}^2 - 2\varphi_i \varphi_{0}\right) - \frac{1}{24}\left(\varphi_i^4 + \varphi_{0}^4 - 4\varphi_i^3\varphi_{0} + 6\varphi_i^2\varphi_{0}^2 - 4\varphi_i\varphi_{0}^3\right)\right].
	\end{align*}
	By collecting terms we can write the full Hamiltonian as
	\begin{align*}
		H=&\sum_{i=0}^{n} \left[ \frac{1}{2}E^C_ip_i^2 + \frac{1}{2}E^J_i\varphi^2_i - \frac{1}{24}E^J_i\varphi_i^4 \right] +  \sum_{i=1}^{n}(K^{-1})_{(i,0)}p_ip_{0} \\
		&+\sum_{i>j = 1}^n(K^{-1})_{(i,j)}p_ip_{j} + \sum_{i=1}^{n}E_{z,i} \left[ - \frac{1}{4}\varphi_i^2\varphi_{0}^2  -\varphi_i\varphi_{0} + \frac{1}{6}\left(\varphi^3_i\varphi_{0} + \varphi_i\varphi_{0}^3\right) \right],
	\end{align*}
	where the effective energy of the capacitances is $E^C_i  = (K^{-1})_{(i,i)}/8$. Note that there is a capacitive coupling between all of the qubits regardless of whether there actually is a capacitor between them. The effective Josephson energies are
	\begin{subequations}
		\begin{align}
			E^J_i =& E_i + E_{z,i} \quad \text{for } i \neq 0,\\
			E^J_{0} =& E_{0} + \sum_{i=1}^n E_{z,i}.
		\end{align}
	\end{subequations}
	We now do the canonical quantization $\varphi_i \rightarrow \hat \varphi_i$ and $p_i \rightarrow \hat p_i$, requiring that $[\hat p_i,\hat \varphi_j] = i\delta_{ij}$. This allows us to change into ladder operators 
	\begin{equation}
	\hat\varphi_i = \sqrt{\frac{\zeta_i}{2}}\left(\hat b_i^\dagger + \hat b_i\right), \qquad \hat p_i = \frac{i}{\sqrt{2\zeta_i}}\left(\hat b_i^\dagger - \hat b_i\right),
	\end{equation}
	with impedance $\zeta_i = \sqrt{(K^{-1})_{(i,i)}/E^J_i}$, and the Hamiltonian takes the form
	\begin{align*}
		\hat H=& \sum_{i=0}^{n} \left[\sqrt{8E^C_iE^J_i} \hat b_i^\dagger \hat b_i - \frac{E^C_i}{12} \left(\hat b_i^\dagger + \hat b_i\right)^4\right] - \frac{1}{2} \sum_{i=1}^{n}\frac{(K^{-1})_{(i,0)}}{\sqrt{\zeta_i\zeta_0}}\left(\hat b_i^\dagger - \hat b_i\right)\left(\hat b_0^\dagger - \hat b_0\right) \\
		& - \frac{1}{2}\sum_{i>j = 1}^n\frac{(K^{-1})_{(i,j)}}{\sqrt{\zeta_i\zeta_j}}\left(\hat b_i^\dagger - \hat b_i\right)\left(\hat b_j^\dagger - \hat b_j\right) \\
		&+\sum_{i=1}^{n} E_{z,i} \left[ - \frac{1}{24}\zeta_i\zeta_0\left(\hat b_i^\dagger + \hat b_i\right)^2\left(\hat b_0^\dagger + \hat b_0\right)^2 -\frac{1}{2}\sqrt{\zeta_i\zeta_0}\left(\hat b_i^\dagger + \hat b_i\right)\left(\hat b_0^\dagger + \hat b_0\right)  \right. \\
		&\phantom{+\sum_{i=1}^{n}}+ \left. \frac{1}{24}\left(\zeta_i\sqrt{\zeta_i\zeta_0}\left(\hat b_i^\dagger + \hat b_i\right)^3\left(\hat b_0^\dagger + \hat b_0\right) + \zeta_0\sqrt{\zeta_i\zeta_0}\left(\hat b_i^\dagger + \hat b_i\right)\left(\hat b_0^\dagger + \hat b_0\right)^3\right) \right].
	\end{align*}
	If the anharmonicities $\alpha_i = E^C_i/2$ of the qubits are sufficiently large, we can justify projecting the Hamiltonian into the lowest two eigenstates of each qubit
	\begin{equation}
	\hat H = -\frac{1}{2} \sum_{i=0}^{n}\omega_i \sigma^z_i +\frac{1}{2}\sum_{n=1}^{n}J^z_i \sigma^z_i \sigma_0^z + \frac{1}{2}\sum_{i=1}^{n}J^x_i \left(\sigma^+_i \sigma_0^-+\sigma^-_i \sigma_0^+\right)   + \frac{1}{2}\sum_{i\neq j = 1}^{n}J^x_{ij}  \left(\sigma^+_i \sigma_j^-+\sigma^-_i \sigma_j^+\right),
	\end{equation}
	where we have neglected terms that do not conserve number excitation, such as $\sigma^\pm_i \sigma^\pm_j$ (this is the rotating wave approximation). We note that the first term is the desired non interacting Hamiltonian, and the second term is the desired Ising coupling term.
	The qubit frequencies and the coupling strengths are given as
	\begin{subequations}\label{eq:params}
		\begin{align}
			\omega_i =& \sqrt{8E^C_iE^J_i} + \frac{1}{2}E^C_i + \frac{1}{6}E_{z,i} \zeta_i \zeta_0   \quad \text {for } i\neq 0, \\
			\omega_{0} =& \sqrt{8E^C_0E^J_0} + \frac{1}{2}E^C_0 + \frac{1}{6}\sum_{i=1}^{n}E_{z,i} \zeta_i \zeta_0, \\
			J_{i}^z =& - \frac{1}{12} E_{z,i} \zeta_i\zeta_0,\\
			J_{i}^x =& \frac{(K^{-1})_{(i,0)}}{\sqrt{\zeta_i\zeta_0}} - E_{z,i}\sqrt{\zeta_i\zeta_0} + \frac{1}{4} E_{z,i}(\zeta_i + \zeta_0) \sqrt{\zeta_i \zeta_0},\label{eq:Jxi0} \\
			J_{ij}^x =& \frac{(K^{-1})_{(i,j)}}{\sqrt{\zeta_i\zeta_j}}.\label{eq:Jxij}
		\end{align}
	\end{subequations}
	If we operate in the weak-coupling regime for the transversal couplings $C_{z,i} \ll C_0,C_i$, for all $i$, the detuning $\delta_{i0} = \omega_i - \omega_0$, of the zeroth qubit compared to all other qubits, becomes much larger than the transverse couplings in \cref{eq:Jxi0}. Using the rotating wave approximation we can then ignore the first-order excitation swaps between these qubits. In this case the Hamiltonian takes the form
	\begin{equation}
	\hat H = \hat H_0 + \frac{1}{2}\sum_{n=1}^{n}J^z_i \sigma^z_i \sigma_0^z + \frac{1}{2}\sum_{i\neq j = 1}^{n}J^x_{ij}  \left(\sigma^+_i \sigma_j^-+\sigma^-_i \sigma_j^+\right).
	\end{equation}
	The last term represents the cross couplings between the $i$th and $j$th qubit for $i,j=1,2,\dots,n$. We get rid of this term as we are in the weak-coupling limit, $C_{z,i} \ll C_{i}$, which makes the Hamiltonian take the desired form.

	\subsection{The driving term}
	
	We are now ready to consider the driving term $\hat H_\text{drive}$. We drive the system capacitively, which yields the driving Lagrangian 
	\begin{equation}
	L_d = \frac{C_d}{2}\left(\dot \varphi_i - \dot\phi_i \right)^2,
	\end{equation}
	which drives the $i$th qubit. We simply add such terms for each qubit we wish to drive. The external driving field is given as
	\begin{equation}
	\phi_i = A \sin(\tilde \Delta_i t + \theta),
	\end{equation}
	where $A$ is the amplitude of the driving, $\tilde \Delta_i$ is the driving frequency, and $\theta$ is the phase. We rewrite the driving terms as
	\begin{align*}
		\phi_i =& A\left(\cos \theta \sin \tilde \Delta_i t + \sin \theta \cos \tilde \Delta_i t\right) ,
	\end{align*}
	where we have expanded the driving field in the in-phase component and the out-of-phase component. Expanding the parentheses of $L_d$ yields
	\begin{equation}
	L_d = \frac{C_d}{2}\left[ \varphi_i^2 + \dot \phi_i^2 - 2\dot\varphi_i\dot \phi_i\right],
	\end{equation}
	where the first term is a kinetic term which can be absorbed into the diagonal of the capacitance matrix, the second term is some irrelevant offset term, and the last term is the interesting term regarding the driving of the system. This alters the conjugate momentum slightly
	\begin{equation}
	\vec p = K\dot{\vec{\varphi}} + \dot{\vec{\phi}},
	\end{equation}
	where $\vec \varphi$ is the vector of node fluxes and $\vec \phi$ is the vector of driving terms. Doing a Legendre transformation the kinetic part of the Hamiltonian takes the form
	\begin{align*}
		H_\text{kin} =& \frac{1}{2}(\vec p - \dot{\vec \phi})^TK^{-1}(\vec p - \dot{\vec \phi}) \\
		=& \frac{1}{2} \left[\vec p^T K^{-1}\vec p + \dot{\vec \phi}^T K^{-1} \dot{\vec \phi} - \vec p^T K^{-1} \dot{\vec \phi} - \dot{\vec \phi}^T K^{-1} \vec p \right].
	\end{align*}
	The first term is the original kinetic term (with the added driving capacitance), the second term is the irrelevant offset term, and the last two terms are the driving terms yielding
	\begin{equation}
	H_d = -\dot \phi_i (K^{-1})_{(i,i)} p_i.
	\end{equation}
	Doing the canonical quantization and changing into step operators we obtain
	\begin{equation}
	\hat{H}_d = -i\dot \phi_i \frac{(K^{-1})_{(i,i)}}{\sqrt{2\zeta_i}}\left(\hat b_i^\dagger - \hat b_i\right).
	\end{equation}
	Truncating to a two-level model as above we find
	\begin{equation}
	\hat H_d = -\dot \phi_i \frac{(K^{-1})_{(i,i)}}{\sqrt{2\zeta_i}} \sigma_i^y,
	\end{equation}
	from which we realize that 
	\begin{equation}
	\beta_i(t) = -A \tilde \Delta_i \frac{(K^{-1})_{(i,i)}}{\sqrt{2\zeta_i}} \left(\cos \theta\cos\tilde \Delta_i t - \sin\theta \sin \tilde \Delta_i t\right),
	\end{equation}
	and if we choose $\theta = 0$ and $\tilde \Delta_i = \omega_i + \Delta_i$ we see that $\Omega = -A\tilde \Delta_i(K^{-1})_{(i,i)}/\sqrt{2\zeta_i} $. Note that we do not necessarily need $\theta=0$ for the gate to work. In fact it can be an advantage to have an out-of-phase component in order to minimize leakage to higher excited states, when the anharmonicity is small using, e.g., pulse-shape engineering schemes \cite{Warren1984,Steffen2003,Rebentrost2009,Motzoi2009,Gambetta2011,deFouquieres2011,Motzoi2013}.

	\section{Realistic parameters}
	
	Here we presents parameters for the circuit model in \cref{fig:CircuitDiagram}(c), which yields the desired gate model of \cref{fig:CircuitDiagram}(a). The parameters are found by calculating the gate model parameters in \cref{eq:params} and then minimizing a cost function which returns a low value when the requirements of the gate model are met. The minimization is done with using the simplex method, with randomized starting points, since many solutions exists. In order to judge the quality of the circuit parameters we also calculate the relative anharmonicity of the two-level systems, i.e., the difference between the 01 and the 12 transition, and the ratio between the effective Josephson energy and the effective capacitive energy.
	
	\begin{table}
		\caption{Circuit and corresponding gate model parameters for possible Toffoli gates. Since the circuit parameter space is rather large we have several possible solutions; some, but far from all, possible solutions are show in the table. The different solutions are labeled and color coded in the first column. The colorcoding corresponds to the simulation results seen in \cref{fig:realFidel}. Column 2-7 show the circuit parameters for the circuit in \cref{fig:CircuitDiagram}(c). Here $E_0$, $E_i$, and $E_{z,i}$ indicate the Josephson junction of the target qubit, the control qubits, and the coupling between them, respectively; $C_0$, $C_i$, and $C_{z,i}$ indicate the capacitance of the target qubit, the control qubits and the coupling between them, respectively. Column 8-12 shows the obtained gate parameters, which can be seen in \cref{eq:params}. Column 13-16 show the quality parameters of the gate: $\alpha_0$ and $\alpha_i$ are the anharmonicities of the target and control qubits, respectively, while $E^J_0/E_0^C$ and $E^J_i/E_i^C$ are the ratios between the effective Josephson energy and effective capacitive energy. The subscript $i$ indicates $i = 1,2$, i.e., the control qubits.}
		\label{tab:realParams}
		\begin{tabular}{c|cccccc|ccccc|cccc}
			\toprule
			\multicolumn{7}{c}{Circuit parameters} & \multicolumn{5}{c}{Gate parameters} & \multicolumn{4}{c}{Quality parameters} \\
			& $E_0$ & $E_i$ & $E_{z,i}$ & $C_0$ & $C_i$ & $C_{z,i}$ & $\omega_0$ & $\omega_i$ & $J^z_i$ & $J^x_i$ & $J^x_{ij} $ & $\alpha_0$ & $\alpha_i$ & $E^{J}_0/E^{C}_0$ & $E_i^J/E_i^C$ \\
			\# & $[2\pi \si{\GHz}]$ & $[2\pi \si{\GHz}]$ & $[2\pi \si{\GHz}]$ & $[\si{\femto\farad}]$ & $[\si{\femto\farad}]$ & $[\si{\femto\farad}]$ & $[2\pi \si{\GHz}]$ & $[2\pi \si{\GHz}]$ & $[2\pi \si{\MHz}]$ & $[2\pi \si{\MHz}]$ & $[2\pi \si{\MHz}]$ & [\%] & [\%] & & \\
			\toprule
			\textcolor[HTML]{00008B}{1}& $20.05$ & $1.26$ & $33.28$ & $15.83$ & $37.84$ & $0.04$ & $30.9$ & $12.8$ & $-320.1$ & $452.6$ & $27.8$ & $-2.0$ & $-2.0$ & $71.1$ & $67.5$ \\ 
			\textcolor[HTML]{0000CD}{2}& $0.61$ & $0.02$ & $29.30$ & $22.19$ & $44.89$ & $0.03$ & $21.9$ & $10.8$ & $-287.2$ & $5.7$ & $14.9$ & $-2.0$ & $-2.0$ & $68.0$ & $68.0$ \\ 
			\textcolor[HTML]{0000FF}{3}& $33.65$ & $1.27$ & $28.70$ & $15.37$ & $43.40$ & $0.03$ & $32.0$ & $11.1$ & $-274.0$ & $453.3$ & $22.5$ & $-2.0$ & $-2.0$ & $72.6$ & $67.2$ \\ 
			\textcolor[HTML]{6A5ACD}{4}& $31.29$ & $0.71$ & $24.10$ & $16.96$ & $53.10$ & $0.03$ & $28.4$ & $9.2$ & $-233.0$ & $254.9$ & $16.7$ & $-2.0$ & $-2.0$ & $69.9$ & $68.0$ \\ 
			\textcolor[HTML]{483D8B}{5}& $4.95$ & $1.20$ & $21.94$ & $27.06$ & $57.03$ & $0.05$ & $17.9$ & $8.5$ & $-214.0$ & $433.7$ & $15.0$ & $-2.0$ & $-2.0$ & $68.5$ & $68.2$ \\ 
			\textcolor[HTML]{7B68EE}{6}& $1.58$ & $1.27$ & $17.14$ & $41.71$ & $56.41$ & $0.09$ & $12.5$ & $7.6$ & $-177.1$ & $474.8$ & $14.6$ & $-1.9$ & $-2.2$ & $77.5$ & $53.7$ \\ 
			\textcolor[HTML]{9370DB}{7}& $39.72$ & $0.11$ & $18.27$ & $17.83$ & $71.69$ & $0.03$ & $26.9$ & $6.8$ & $-176.0$ & $38.3$ & $9.8$ & $-2.0$ & $-2.0$ & $70.4$ & $68.0$ \\ 
			\textcolor[HTML]{663399}{8}& $0.56$ & $0.03$ & $17.53$ & $35.36$ & $77.67$ & $0.08$ & $13.4$ & $6.4$ & $-172.2$ & $12.4$ & $13.1$ & $-2.0$ & $-2.0$ & $65.3$ & $70.5$ \\ 
			\textcolor[HTML]{8A2BE2}{9}& $45.01$ & $1.22$ & $16.13$ & $17.72$ & $76.22$ & $0.02$ & $27.1$ & $6.4$ & $-154.6$ & $437.1$ & $7.8$ & $-2.0$ & $-2.0$ & $70.9$ & $68.3$ \\ 
			\textcolor[HTML]{4B0082}{10}& $19.87$ & $1.22$ & $11.35$ & $31.88$ & $105.46$ & $0.04$ & $15.1$ & $4.6$ & $-109.1$ & $437.3$ & $4.8$ & $-2.0$ & $-2.0$ & $70.2$ & $68.4$ \\ 
			\textcolor[HTML]{9932CC}{11}& $57.40$ & $1.25$ & $7.45$ & $19.24$ & $153.26$ & $0.01$ & $24.9$ & $3.2$ & $-70.6$ & $444.1$ & $1.7$ & $-2.0$ & $-2.0$ & $71.9$ & $68.8$ \\ 
			\textcolor[HTML]{9400D3}{12}& $23.81$ & $0.80$ & $6.30$ & $37.75$ & $187.62$ & $0.02$ & $12.7$ & $2.6$ & $-60.1$ & $287.2$ & $1.0$ & $-2.0$ & $-2.0$ & $71.0$ & $68.9$ \\ 
			\textcolor[HTML]{BA55D3}{13}& $58.41$ & $0.09$ & $3.85$ & $21.01$ & $338.06$ & $0.00$ & $22.7$ & $1.4$ & $-36.6$ & $31.3$ & $0.0$ & $-2.0$ & $-2.0$ & $71.7$ & $68.7$ \\ 
			\textcolor[HTML]{D8BFD8}{14}& $0.01$ & $0.19$ & $2.50$ & $299.11$ & $386.88$ & $0.37$ & $1.7$ & $1.1$ & $-25.8$ & $71.1$ & $1.3$ & $-1.9$ & $-2.2$ & $77.5$ & $53.7$ \\  
			\toprule
		\end{tabular}
	\end{table}
	
	In order to simplify the numerical investigation we have assumed that the parameters of the control qubits are identical. The parameters obtained are presented in \cref{tab:realParams}. As expected we see that the capacitance of the coupling $C_{z,i}$ should be low compared to the other couplings as we wish to operate in the weak coupling regime. We note that we get Ising couplings in the range $|J^z| \in [25,320]$ and in all cases dominating the cross coupling $J^x_{ij}$. The swapping couplings $J^x_i$ are all several factors lower than the detunings $\Delta_{i0} = |\Omega_i - \Omega_0|$.
	
\begin{figure}
		\centering
		\includegraphics[scale=0.5]{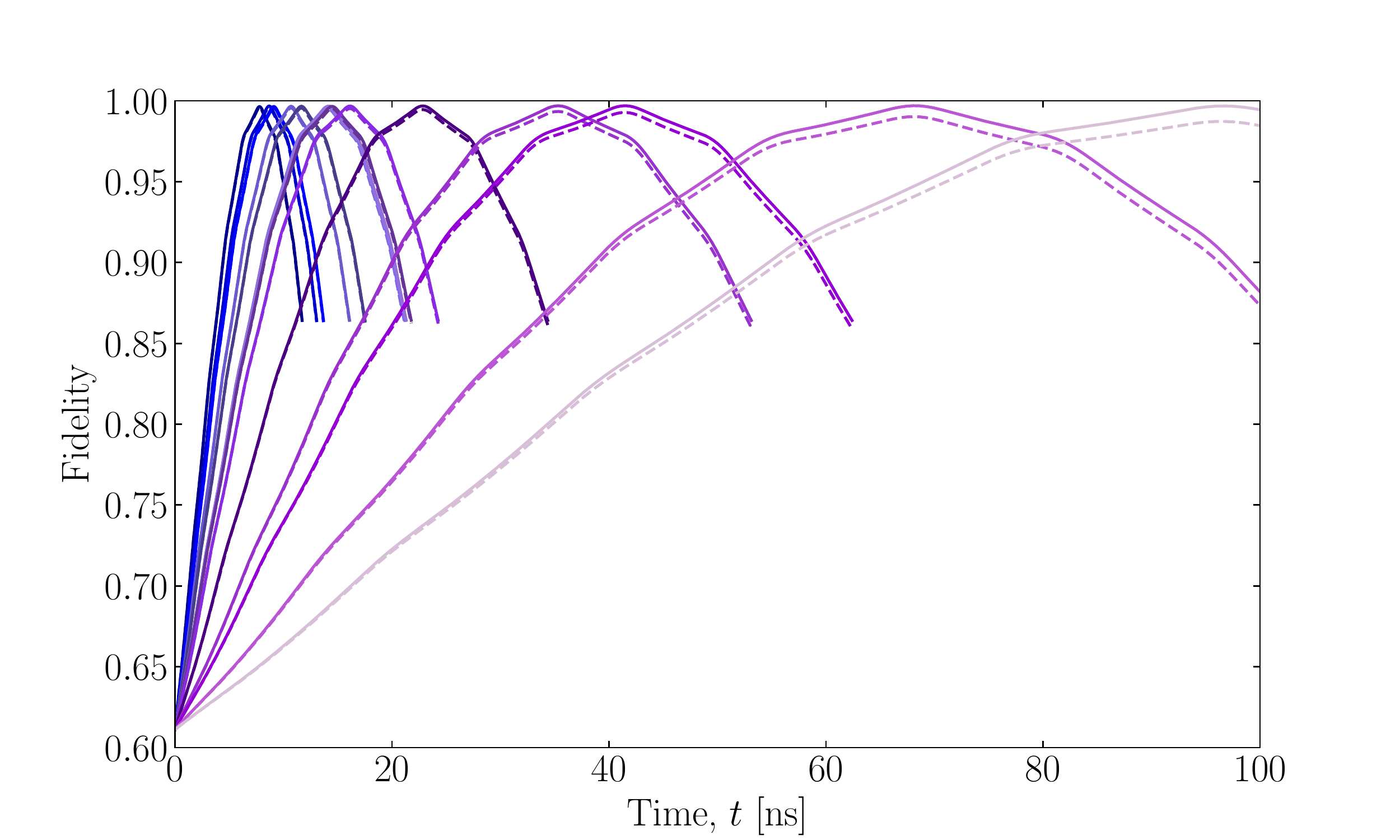}
		\caption{Average fidelity as a function of time for all the gate configurations presented in \cref{tab:realParams}. All simulations are done at a driving of $\Omega=J^z/8$. The fidelity is expected to peak at $T=\pi/2\Omega$ The solid lines are simulations without decoherence, while the dashed line includes decoherence. The gate with the lowest gate time corresponds to gate \#1 in \cref{tab:realParams} and so fourth. The color of the lines also corresponds to the colors in \cref{tab:realParams}.}
		\label{fig:realFidel}
	\end{figure}

	We simulate all of the gates in \cref{tab:realParams} and find that all result in a maximum fidelity above 0.99, when the driving is $\Omega=J^z_i/8$. The average fidelity as a function of time can be seen in \cref{fig:realFidel}.
	
	\section{Analytical treatment of the operator norm} \label{app:opnorm}
\begin{figure}
		\centering
		\includegraphics[width=.5\columnwidth]{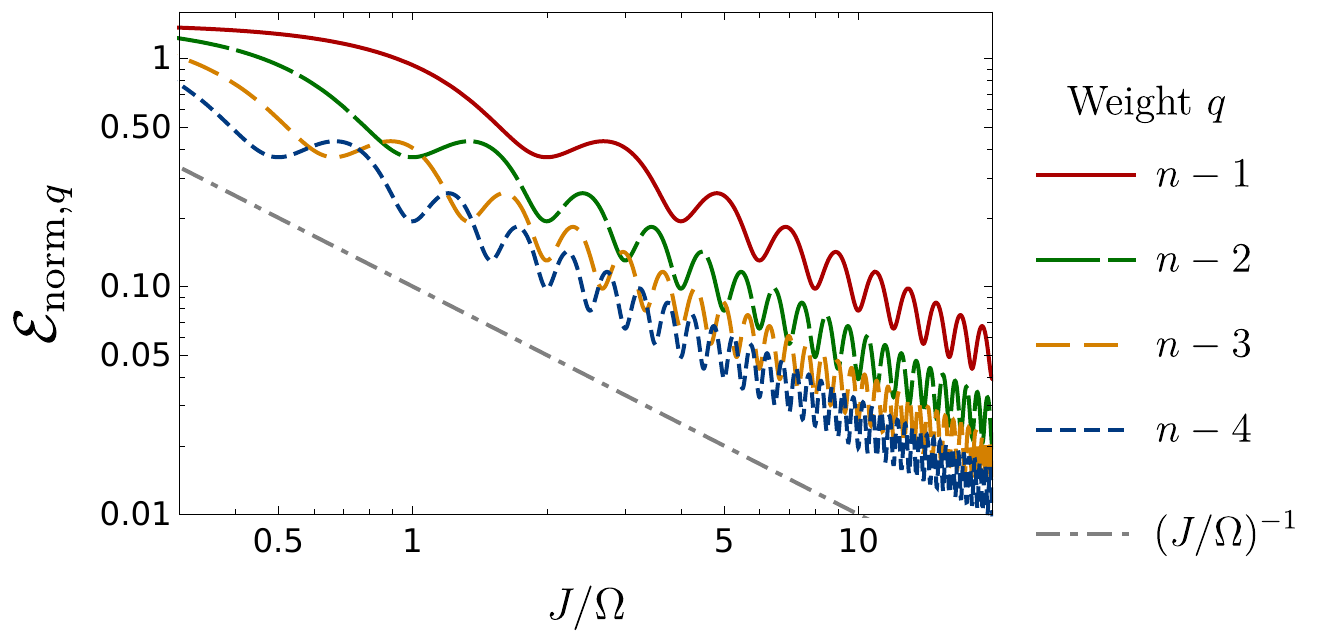}
		\caption{Operator norm error $\mathcal{E}_{\text{norm},q}$ contributions due to subspaces with weights $n-q=1, \ldots 4$, at various protocol times. The overall error $\mathcal{E}_\text{norm}$ for the whole gate is always the maximum and hence is completely determined by $q=n-1$. }
		\label{fig:Enorm}
	\end{figure}

	The operator norm of the difference of two matrices is given by 
	\begin{align*}
		\mathcal{E}_\text{norm}(U, U_\text{goal}) = || U - U_\text{goal} ||
	\end{align*}
	which returns the largest eigenvalue of $U - U_\text{goal}$ and hence captures the worst-case error of our gate. In the case of permutation symmetry among all work qubits, the expression simplifies to
	\begin{align*}
		\mathcal{E}_\text{norm}(U, U_\text{goal}) &= \max_q || U_q - U_\text{goal,q} ||.    
	\end{align*}
	For the case of the resonant Toffoli gate, we obtain the following:
	\begin{align}
		\hat{ U }_q(T) - \hat{ U }_{\text{goal},q}(T) 
		=& \ \unity_2 \Big[ -\cos \left( \frac{\pi \gamma}{2} \right) + \cos \left( \frac{\pi}{2} \sqrt{1+\gamma^2} \right) \Big] \\
		&+ i \sigma^z \Big[ \sin \left( \frac{\pi \gamma }{2} \right) - \frac{\gamma}{\sqrt{1+\gamma^2}} \sin \left( \frac{\pi}{2} \sqrt{1+\gamma^2} \right)    \Big] \\
		&  - i(\cos(\theta) \sigma^x + \sin(\theta) \sigma^y ) \frac{ \sin \left( \frac{\pi}{2} \sqrt{1+\gamma^2} \right) }{ \sqrt{1+\gamma^2} }
	\end{align}
	From this expression we can  efficiently calculate the exact operator norm error in each individual weight-$q$ subspace. The results are shown in \cref{fig:Enorm}. It is clear that the most resonant subspace, with $q=n-1$, always contributes the largest error. Therefore, the $\max$ operation can be dropped in $\mathcal{E}_\text{norm}$, and we find that operator norm error is actually independent of $n$. All in all, we find that 
	\begin{align}
		\mathcal{E}_\text{norm}^2 = 2 - 2 \cos \left( \frac{J \pi}{2 \Omega} \right) \cos \left( \frac{\pi}{2} \sqrt{1+\frac{J^2}{\Omega^2}}  \right) 
		- 2 \frac{J/\Omega}{\sqrt{1+\frac{J^2}{\Omega^2}} } \sin \left( \frac{J \pi}{2 \Omega} \right) \sin \left( \frac{\pi}{2} \sqrt{1+\frac{J^2}{\Omega^2}} \right).
	\end{align}

	\section{Steane encoding states}\label{app:steane}
	
	We derive the two Steane encoding states $|0\rangle_L$ and $|1\rangle_L$, by applying the \textsc{cnot}$^n$ gates to the initial states
	\begin{align*}
		|+++0000 \rangle =& \frac{1}{2\sqrt{2}} (|0\rangle + |1\rangle)(|0\rangle + |1\rangle)(|0\rangle + |1\rangle)|0000\rangle \\
		\xrightarrow{\text{CNOT}^2} & \frac{1}{2\sqrt{2}} (|0\rangle + |1\rangle)(|0\rangle + |1\rangle)(|0\rangle + |1\rangle)|0000\rangle \\
		\xrightarrow{\text{CNOT}^3} & \frac{1}{2\sqrt{2}} (|0\rangle + |1\rangle)(|0\rangle + |1\rangle ) \left(|00000\rangle + i |11101\rangle \right)\\
		\xrightarrow{\text{CNOT}^3} & \frac{1}{2\sqrt{2}} (|0\rangle + |1\rangle) \left(|000000\rangle + i |101011\rangle + i (|011101\rangle + i |110110\rangle) \right)\\
		\xrightarrow{\text{CNOT}^3} & |0\rangle_L =  \frac{1}{2\sqrt{2}} (|0000000\rangle + i |0101011\rangle +i|0011101\rangle - |0110110\rangle \\ & \phantom{|0\rangle_L = \frac{1}{2\sqrt{2}}(} + i |1000111\rangle - |1101100\rangle - |1011010\rangle -i|1110001\rangle ),
	\end{align*}
	\begin{align*}
		|+++1000 \rangle =& \frac{1}{2\sqrt{2}} (|0\rangle + |1\rangle)(|0\rangle + |1\rangle)(|0\rangle + |1\rangle)|1000\rangle \\
		\xrightarrow{\text{CNOT}^2} & \frac{1}{2\sqrt{2}} (|0\rangle + |1\rangle)(|0\rangle + |1\rangle)(|0\rangle + |1\rangle)|1110\rangle \\
		\xrightarrow{\text{CNOT}^3} & \frac{1}{2\sqrt{2}} (|0\rangle + |1\rangle)(|0\rangle + |1\rangle ) \left(|01110\rangle + i |10011\rangle \right)\\
		\xrightarrow{\text{CNOT}^3} & \frac{1}{2\sqrt{2}} (|0\rangle + |1\rangle) \left(|001110\rangle + i |010011\rangle + i (|100101\rangle + i |111000\rangle) \right)\\
		\xrightarrow{\text{CNOT}^3} & |1\rangle_L =  \frac{1}{2\sqrt{2}} (-|0001110\rangle - i |0010011\rangle -i|0100101\rangle + |0111000\rangle \\ & \phantom{|1\rangle_L = \frac{1}{2\sqrt{2}}(} - i |1001001\rangle + |1100010\rangle + |1010100\rangle + i|1111111\rangle ).
	\end{align*}
	
\end{document}